\documentclass[runningheads]{llncs}
\usepackage{graphicx}
\usepackage{subcaption}
\usepackage{hyperref}
\usepackage{tikz}
\usepackage{comment}
\usepackage{amsmath,amssymb} % define this before the line numbering.
\usepackage{color}

\usepackage[accsupp]{axessibility}  % Improves PDF readability for those with disabilities.

\begin{document}
% \linenumbers
\pagestyle{headings}
\mainmatter
\def\ECCVSubNumber{6817}  % Insert your submission number here

\title{Expanded Adaptive Scaling Normalization for\\ End to End Image Compression} % Replace with your title

\begin{comment}
\titlerunning{ECCV-22 submission ID \ECCVSubNumber} 
\authorrunning{ECCV-22 submission ID \ECCVSubNumber} 
\author{Anonymous ECCV submission}
\institute{Paper ID \ECCVSubNumber}
\end{comment}
%******************

%\begin{comment}
\titlerunning{EASN}
\author{Chajin Shin\index{Shin, Chajin} \and
Hyeongmin Lee \and
Hanbin Son \and
Sangjin Lee \and
Dogyoon Lee\and
Sangyoun Lee
}
\authorrunning{C. Shin et al.}
\institute{School of Electrical and Electronic Engineering, Yonsei University, Seoul, Korea\\
$\{chajin,minimonia,hbson,pandatimo,nemotio,syleee\}@yonsei.ac.kr$}
%\end{comment}
%******************

\maketitle

\begin{abstract}
Recently, learning-based image compression methods that utilize convolutional neural layers have been developed rapidly. Rescaling modules such as batch normalization which are often used in convolutional neural networks do not operate adaptively for the various inputs. Therefore, Generalized Divisible Normalization(GDN) has been widely used in image compression to rescale the input features adaptively across both spatial and channel axes.
However, the representation power or degree of freedom of GDN is severely limited. Additionally, GDN cannot consider the spatial correlation of an image. To handle the limitations of GDN, we construct an expanded form of the adaptive scaling module, named Expanded Adaptive Scaling Normalization(EASN). First, we exploit the swish function to increase the representation ability. Then, we increase the receptive field to make the adaptive rescaling module consider the spatial correlation. Furthermore, we introduce an input mapping function to give the module a higher degree of freedom. We demonstrate how our EASN works in an image compression network using the visualization results of the feature map, and we conduct extensive experiments to show that our EASN increases the rate-distortion performance remarkably, and even outperforms the VVC intra at a high bit rate.

\end{abstract}

\section{Introduction}
\vspace{-0.2cm}
Image compression is one of the most important and fundamental tasks in image processing and computer vision. There are a countless digital images in the world, and numerous new images are generated every day. Therefore, image compression is essential to save and transmit these massive images efficiently. Many classic image compression codecs have been developed, including JPEG~\cite{JPEG}, JPEG2000~\cite{JPEG2000}, HEVC~\cite{hevc}, and VVC~\cite{vvc}. They use several classic methods such as transformation, quantization, and entropy coding to reduce redundant information in image.

\begin{figure}
\centering
\begin{subfigure}{.49\textwidth}
    \centering
    \includegraphics[width=.95\linewidth]{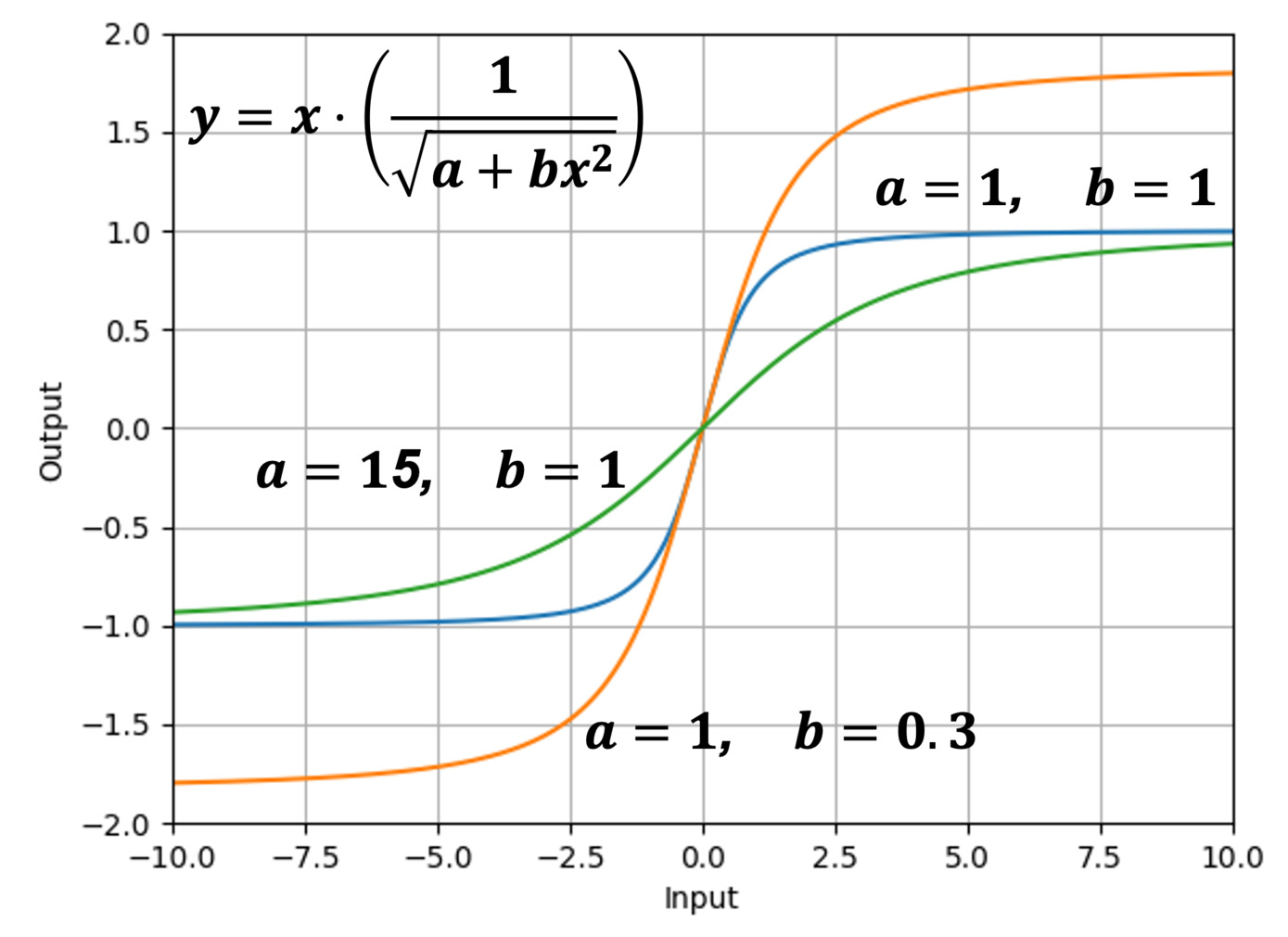}  
    \caption{Simple version of GDN~\cite{end-to-end}}
    \label{fig:simple_GDN}
\end{subfigure}
\begin{subfigure}{.49\textwidth}
    \centering
    \includegraphics[width=.95\linewidth]{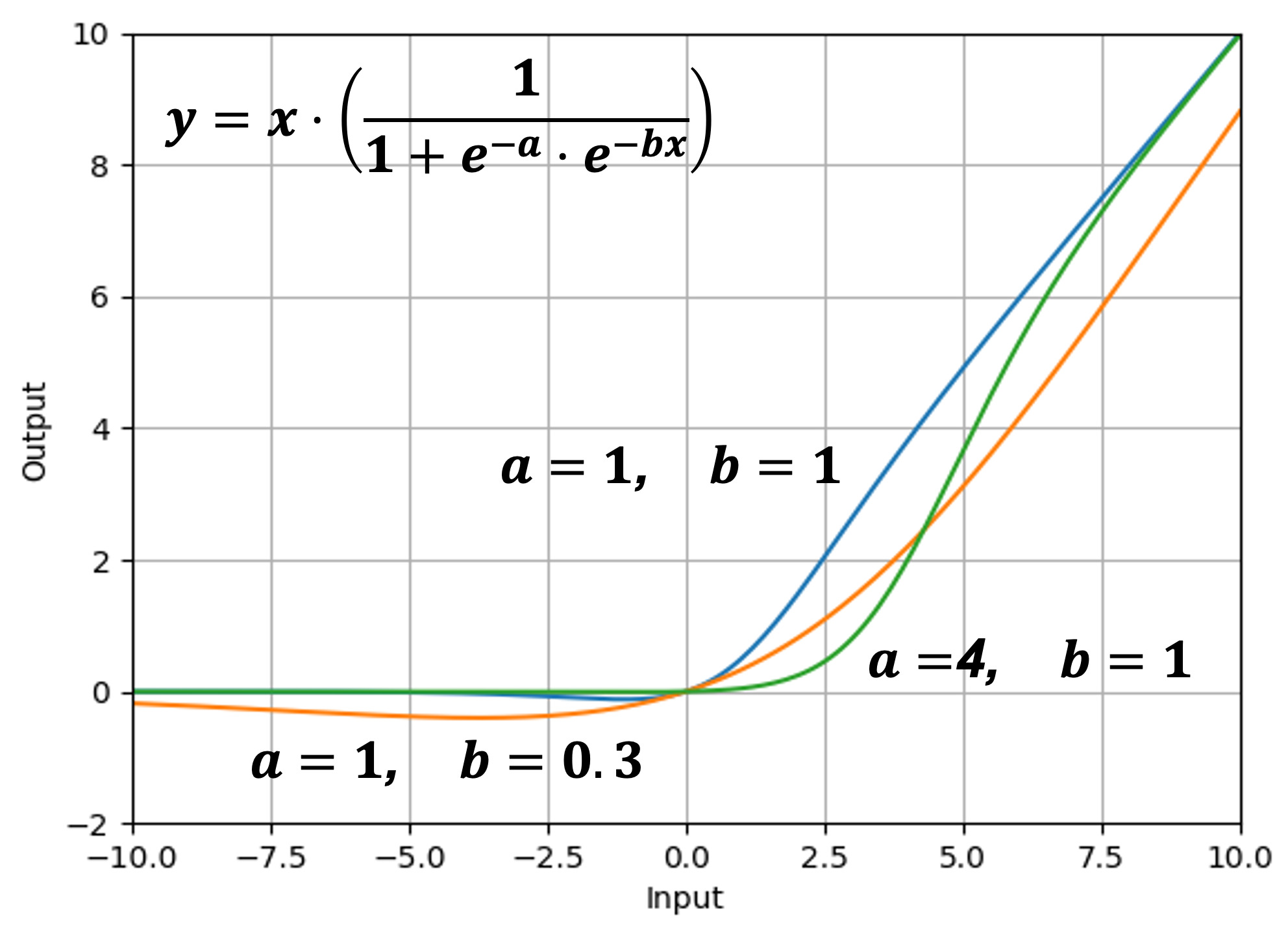}
    \caption{Simple version of Ours}
    \label{fig:simgle_Ours}
\end{subfigure}
\caption{Comparison between simple version of GDN~\cite{end-to-end} and ours. $a$ and $b$ are the learnable scalar parameters. We exploit the swish~\cite{swish} function (b) instead of the arithmetic sigmoidal function (a) from GDN~\cite{end-to-end}.}
\label{fig:simple_GDN_Ours_comparison}
\vspace{-0.2cm}
\end{figure}

Recently, deep learning-based image processing techniques have emerged that have shown superior performance in many computer-vision tasks. There have been many attempts to apply deep learning-based methods to image compression, and convolutional Variational Autoencoder(VAE)-based architecture, which has an hourglass-shape with the encoder and decoder, is the mainstream in image compression. Ball{\'e} \textit{et al}.~\cite{end-to-end} propose a differentiable method for both quantization and bit estimation, and entropy can be optimized directly using this method effectively. Then, HyperPrior~\cite{hyperprior}, which uses additional bits to model the latent vectors as Gaussian distribution, has been proposed to further reduce redundancy in latent vectors. Since then, various models have been proposed~\cite{gmm,asymmetric,checkerboard,context,joint,channel-wise,end-to-end-attention}, and many models now compete with VVC intra of the traditional codec.

Many methods that use a convolutional VAE structure~\cite{end-to-end,hyperprior,asymmetric,checkerboard,context,joint,channel-wise,end-to-end-attention} utilize Generalized Divisible Normalization (GDN)~\cite{end-to-end} instead of using both existing rescaling modules and activation functions, such as batch normalization~\cite{batchnorm} and ReLU~\cite{relu} function. The reason is that existing rescaling modules and activation functions cannot operate adaptively for the various inputs since they apply the same value or manner to all spatial locations equally. In contrast, GDN controls the input value scale of intermediate features adaptively and non-linearly across spatial and channel axes.

However, GDN has several limitations.
First, the representation power of GDN is severely limited. The reason is that GDN can only have non-negative learnable parameters, and the input features of GDN should be squared since they are included within the square root of GDN. We empirically find that increasing the receptive field of GDN or adding more layers cannot increase the performance due to the square root term in the equation of GDN.
Second, GDN has only a $1\times1$ receptive field, which cannot deal with the spatial correlation of images. Natural images have strong redundancy between adjacent pixels, thus spatial correlation must be considered for image compression.
Third, GDN has a limitation in non-linearity because it has a only single $1\times1$ convolutional layer.
Finally, GDN is unstable at the training stage when we use a larger convolutional kernel or add more layers. Although GDN is initialized as a scaled identity matrix for convergence, it is insufficient to stabilize the adaptive rescaling module.

In this paper, we propose Expanded Adaptive Scaling Normalization(EASN), which is an expanded form of the adaptive scaling module, to overcome the limitations of GDN.
First, we exploit the swish~\cite{swish} function instead of the sigmoidal function of GDN, as shown in Fig.~\ref{fig:simple_GDN_Ours_comparison}. Since the swish function has no square root, both the non-negative and negative learnable parameters are available, and the input feature does not need to be squared. This allows the scaling module to utilize the full range of parameters and inputs.
Second, we increase the receptive field and add more layers, which allows the scaling module to consider the spatial correlation of the features and approximate more complex functions, and utilize the skip connection to stabilize the training.
Additionally, we add an input mapping function to the scaling module to transform the input features to increase the degree of freedom of modules.
Furthermore, we use the features before the spatial resolution is reduced by downsampling for a scaling function, and obtain a better performance.
Moreover, we show that simply increasing the layers of EASN does not increase the performance by ablation study, and we propose a structure that makes the EASN deeper effectively to further improve performance.
Finally, we visualize the output feature map of the scaling function of low and high bit rate models and reveal that our EASN can adjust and scale the highfrequency components in accordance with the bit rate. We evaluate our model on the Kodak dataset~\cite{kodak} and CLIC2021 validation dataset~\cite{clic}, and our EASN achieves the rate-distortion performance dramatically, even outperforms VVC intra at a high bit rate.

\section{Related Works}
\vspace{-0.2cm}
\textbf{Traditional Codec}.
There are various traditional hand-crafted image compression methods. JPEG~\cite{JPEG}, JPEG2000~\cite{JPEG2000}, HEVC~\cite{hevc} and VVC~\cite{vvc} are very popular image compression standard methods. To compress and reduce spatial redundancy of the image effectively, encoder modules divide the image into multiple blocks, and convert spatial domain of the image to the frequency domain with traditional transforms such as discrete cosine transform(DCT). After transforming, quantization and entropy coding, like Huffman coding, are conducted. Moreover, HEVC or VVC have many modes of each module and they check every case to get best rate-distortion performance.

\noindent
\textbf{Learning-based}.
Recently, deep learning-based image processing methods have emerged and shown superior performance in various computer vision tasks, and there have been many efforts to utilize deep learning-based methods for image compression. In the first stage, some works~\cite{toderici2015variable,toderici2017full} utilize recurrent neural networks for image compression. These methods can have variable bit rates using the recurrent scheme. However, entropy of the image is not optimized directly since the constraint of entropy is not in the loss function, thus these methods show lower performance than JPEG2000.

The second stage, which is convolutional VAE-based architectures, has become the mainstream in image compression with optimizing entropy directly through loss function. From Ball{\'e} \textit{et al}.~\cite{end-to-end}, minimizing the expectation of Kullback-Leibler divergence is equal to minimizing distortion and entropy at the same time using a variational autoencoder. Furthermore, they~\cite{end-to-end} proposes a differentiable method for both quantization and bit estimation to consider the bit rate constraint at the training step.
They add uniform random noise in the range of $[-0.5, 0.5]$ to the latent representation $y$, which is the output of encoder $g_a$. By adding the noise, they can approximate the probability mass function (PMF) of the quantized latent representation $\hat{y}$ with integrating the probability density function (PDF) of latent representation $y$. Using approximated PMF, they~\cite{end-to-end} directly optimize the entropy of the image with the following total loss function.

\begin{align}
\vspace{-5mm}
  \mathcal{L}=-\mathbb{E}[log_2P]+\lambda \cdot D(x,\hat{x})
\end{align}

\noindent
where $P$ is the estimated PMF of the latent representation, and $D$ is the distortion between the original image $x$ and the reconstruction $\hat{x}$. Thereafter, Hyperprior~\cite{hyperprior} introduces an auxiliary convolutional autoencoder to utilize side information to model the latent representation $y$ as a Gaussian distribution to further reduce the spatial redundancy in the latent representation $y$. Through these, the performance of convolutional VAE-based image compression has greatly improved. However they~\cite{end-to-end,hyperprior} are still transform-based models, and there are no spatial or context prediction modules. Some works~\cite{context,joint} predict the context of an image by using an autoregressive context prediction module with latent representation $y$ and Hyperprior. Another work~\cite{checkerboard} proposes a parallelizable context model to accelerate the sequential process of the autoregressive context prediction module. Further works~\cite{gmm,asymmetric} consider the latent representation $y$ as a more generalized distribution such as the asymmetric Gaussian or Gaussian mixture distribution.

\section{Preliminary}
\vspace{-0.2cm}
Existing rescaling modules and activation functions such as batch normalization~\cite{batchnorm} or ReLU~\cite{relu} function are not adaptive since they operate the same way to all spatial location equally. To deal with this problem, Ball{\'e} \textit{et al}.~\cite{end-to-end} propose GDN, which rescale input features adaptively and non-linearly across spatial and channel axes. GDN is used in image compression neural network instead of batch normalization~\cite{batchnorm} or ReLU~\cite{relu} functions. GDN of normal version $g_i$ is used in the analysis transform, which is encoder, and inverse version $g_i^{inv}$ is used in the synthesis transform, which is decoder.

\begin{align}
\vspace{-2mm}
  \label{eqn:normal_gdn}
  g_i(m,n) &= x_i(m,n)\cdot\frac{1}{\sqrt{\beta_i+\sum_j\gamma_{ij}(x_j(m,n))^2}}\\
  \label{eqn:inv_gdn}
  g_i^{inv}(m,n) &= x_i(m,n)\cdot\sqrt{\beta_i+\sum_j\gamma_{ij}(x_j(m,n))^2)}
\end{align}

\noindent
 where $i$ is output channel index, and $j$ is input channel index. We can interpret $\gamma$ as the $1\times1$ convolutional kernel, and $\beta$ as the bias. $(m, n)$ are coordinates of spatial height and width axis. If we focus on the normal version of GDN, $g_i$, it can be simplified as follows.

\begin{align}
  g=\frac{x}{\sqrt{a+bx^2}}
\end{align}

\noindent
where $a$ and $b$ are non-negative scalar learnable parameters. Output $g$ is adaptively changed according to input $x$, since scaling factor function, which is rescaling part $s(x) = 1 / \sqrt{a+bx^2}$, is various with respect to input features $x$. Therefore, we can consider GDN as the adaptive rescaling module. Furthermore, Fig.~\ref{fig:simple_GDN} shows graph of output $y$ with respect to input $x$ with different values of $a$ and $b$. We can find that the network can learn non-linear sigmoidal shape using $a$ and $b$, and can use it as a learnable sigmoidal shape activation function. GDN uses multivariate parameters for $a$ and $b$ instead of the scalar parameter, thus GDN is a multivariate sigmoidal function.

\section{Method}
\vspace{-0.2cm}
In this section, we introduce the limitations of GDN~\cite{end-to-end} and our proposals to cope with the limitations. Furthermore, we propose more deeper scaling module architecture to obtain higher performance.

\begin{figure}[t]
\centering
\begin{subfigure}{.49\textwidth}
    \centering
    \includegraphics[width=.98\linewidth]{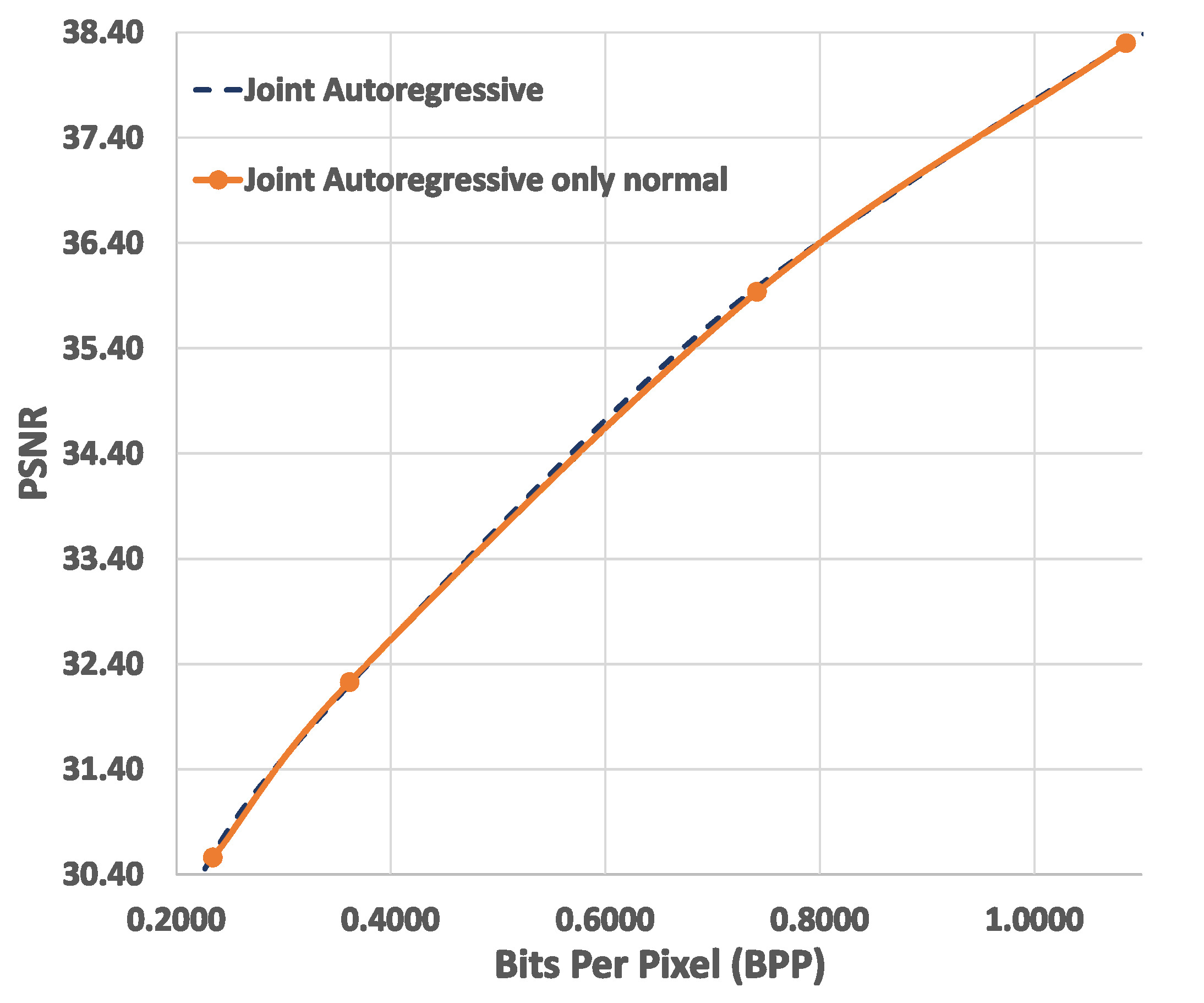}  
    \caption{JA~\cite{joint} with only normal GDN~\cite{end-to-end}}
    \label{fig:JA_reverse}
\end{subfigure}
\begin{subfigure}{.49\textwidth}
    \centering
    \includegraphics[width=.98\linewidth]{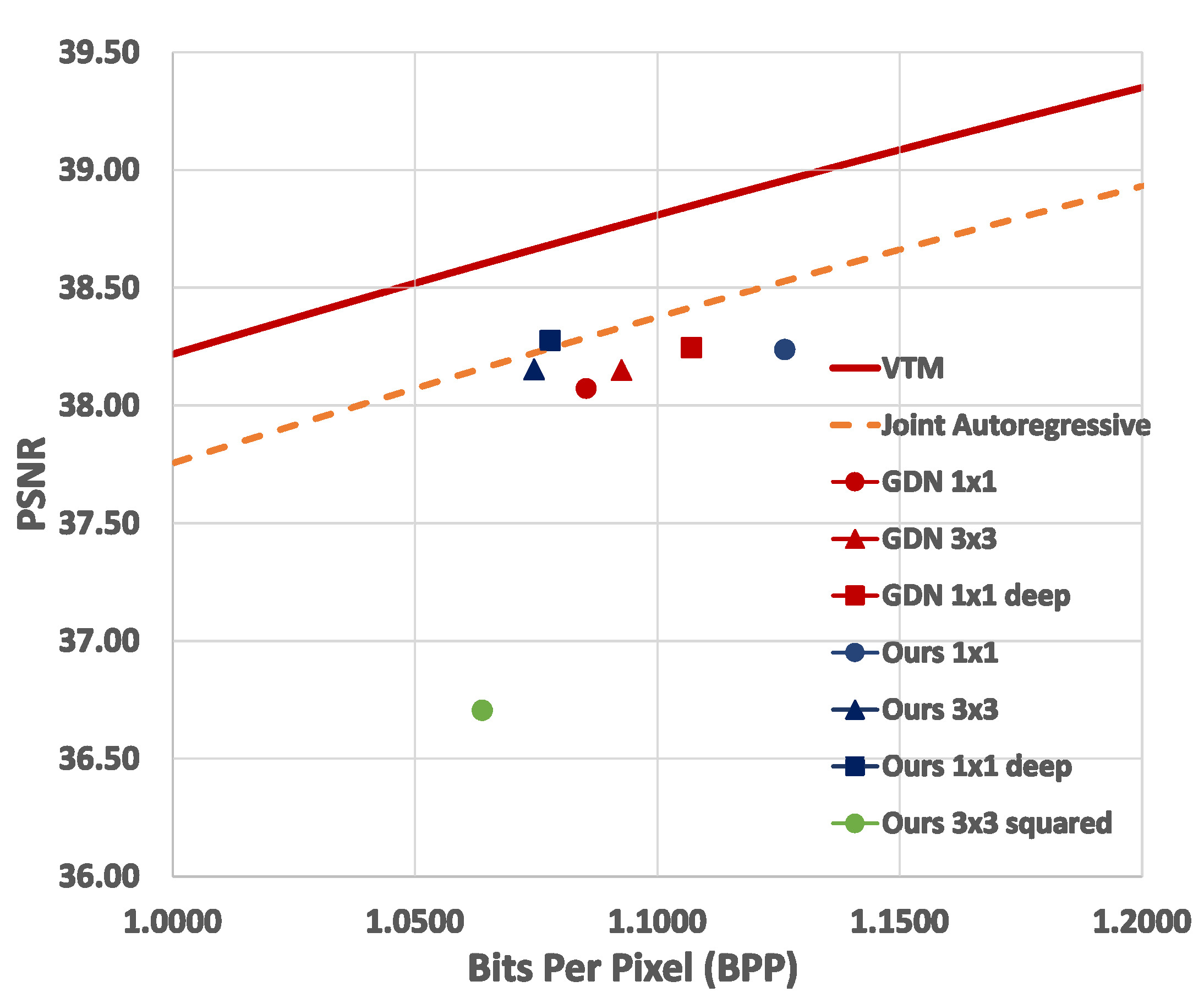}  
    \caption{Representation power limitation}
    \label{fig:GDN_limitation}
\end{subfigure}
\caption{(a): There is no difference in performance even if only the normal version of GDN~\cite{end-to-end} is used. (b): Red points denote Eq.~\ref{eqn:scaling_factor_function_of_GDN} of GDN, and blue points indicate Eq.~\ref{eqn:scaling_factor_function_of_ours} of ours. The performance increases steadily with Eq.~\ref{eqn:scaling_factor_function_of_ours} of ours. By contrast, Eq.~\ref{eqn:scaling_factor_function_of_GDN} of GDN does not show a performance increase owing to limitations of representation power or degree of freedom.}
\vspace{-0.2cm}
\end{figure}

The scaling module of GDN can be expressed as follows.

\begin{align}
\vspace{-0.2cm}
  g(x)=x\cdot s(x)
  \label{eqn:gdn_representation}
\end{align}

\noindent
where $x$ is an input feature and $s(\cdot)$ is a scaling factor function. We only consider the normal version of GDN $g_i$ in Eq.~\ref{eqn:normal_gdn} for scaling factor function $s(x)$ in this section. We replace the inverse version of GDN $g_i^{inv}$ in Eq.~\ref{eqn:inv_gdn} with the normal version $g_i$ to consider only a single case when we modify the scaling factor function $s(x)$ in GDN. We empirically confirm that the network shows the same rate-distortion performance when we only use the normal version of GDN. Fig.~\ref{fig:JA_reverse} represents the result that Joint Autoregressive~\cite{joint} model with only normal version of GDN shows the same performance as the base Joint Autoregressive model. Therefore, we only use the normal version for simplicity.

\subsection{Swish Function}\label{sec:method-swish_function}
\vspace{-0.2cm}
In this section, we describe that GDN~\cite{end-to-end} has limited representation power or degree of freedom, and we show that using a swish~\cite{swish} function for an adaptive rescaling module allows it to cope with the problem of GDN.

Considering the scaling factor function $s(x)$ of GDN,

\begin{align}
\vspace{-0.2cm}
  s(x) = \frac{1}{\sqrt{\beta_i+\sum_j\gamma_{ij}(x_j(m,n))^2}}
\label{eqn:scaling_factor_function_of_GDN}
\end{align}

\noindent
we can notice that $\beta_i+\sum_j\gamma_{ij}(x_j(m,n))^2$ should be non-negative. First, to keep it non-negative, GDN set $\beta_i$ and $\gamma_{ij}$ as non-negative learnable parameters, which limits the degree of freedom of the rescaling module. Second, the input features $x$ should be squared to be non-negative. This leads to information loss because the two different values that have the same magnitude but opposite sign attain the same value after the square operation. Finally, the scaling factor function of Eq.~\ref{eqn:scaling_factor_function_of_GDN} is even symmetric, and it equally scales for inputs that have the same magnitude but opposite sign. These characteristics significantly limit the representation power and degree of freedom of the rescaling module. Therefore, we modify the scaling factor function $s(\cdot)$.

In Eq.~\ref{eqn:scaling_factor_function_of_GDN}, $\beta_i$ is a vector with output channel axis $i$, and the convolution operation $\sum_j\gamma_{ij}x_j^2$ is calculated along the input channel axis $j$. Thus, we can factorize Eq.~\ref{eqn:scaling_factor_function_of_GDN} by $\frac{1}{\sqrt{\beta_i}}$ with $\delta_{ij}= \frac{\gamma_{ij}}{\beta_i}$. Then, $\frac{1}{\sqrt{\beta_i}}$ can be considered a constant scaling factor along the input channel axis of the next convolutional layer. This means that the convolution kernel of next layer can learn a constant scaling factor $\frac{1}{\sqrt{\beta_i}}$, thus we can ignore this term. Therefore, we can consider the scaling factor function as below.

\begin{align}
\vspace{-0.2cm}
  \bar{s}(x) = \frac{1}{\sqrt{1+\sum_j\delta_{ij}(x_j(m,n))^2}}
  \label{eqn:normalized_scaling_factor_fucntion_of_gdn}
\end{align}

\noindent
We replace the even symmetric function of Eq.~\ref{eqn:normalized_scaling_factor_fucntion_of_gdn} with a sigmoid function that has the same output range of $[0, 1]$, but is a bijective function as follows.

\begin{figure}[t]
\centering
\begin{subfigure}{.93\textwidth}
    \centering
    \includegraphics[width=.99\linewidth]{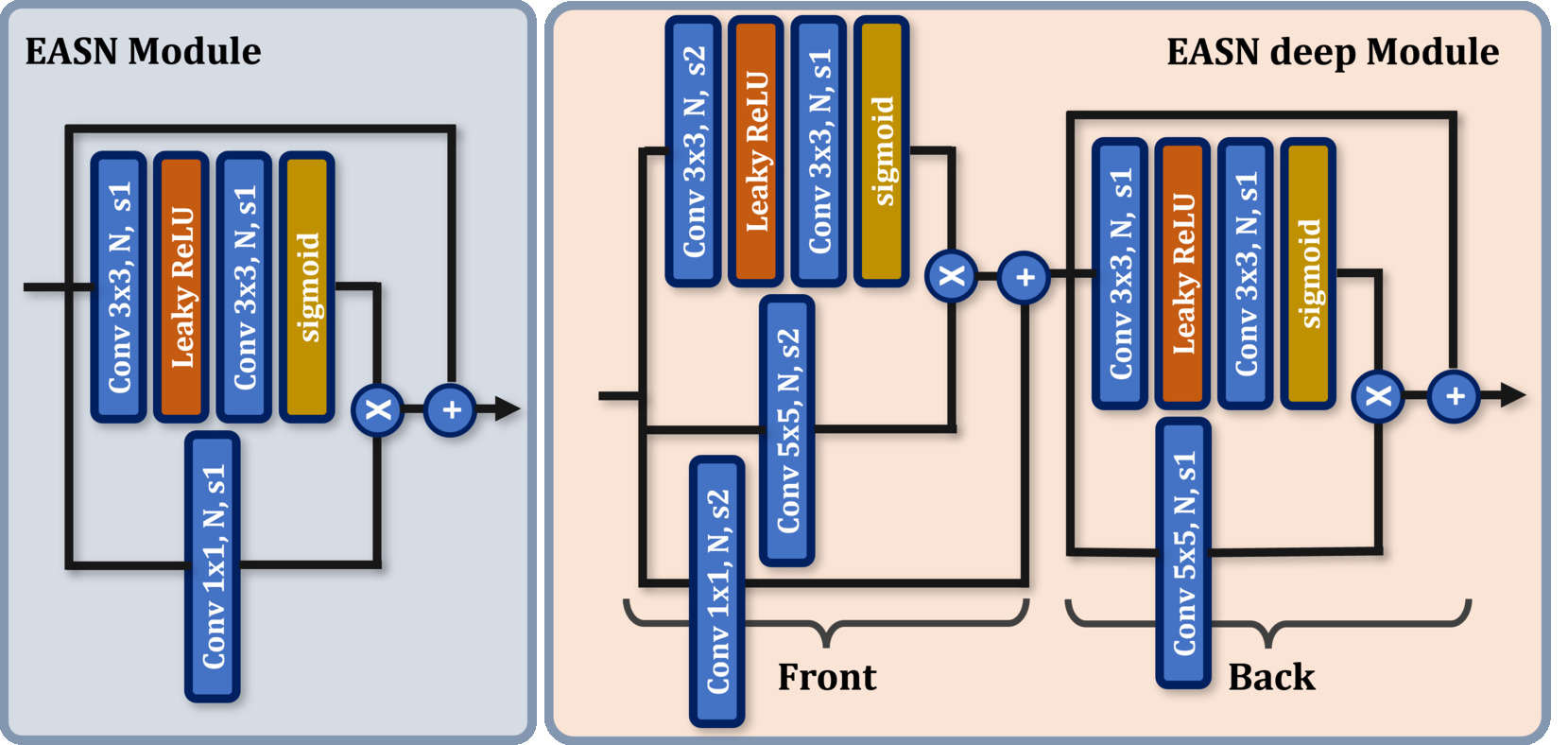}
\end{subfigure}
\caption{Our EASN and EASN-deep. $N$ is output channel, and $s1$, $s2$ represent stride 1 and 2, respectively.}
\label{fig:modules}
\vspace{-0.2cm}
\end{figure}

\begin{align}
\vspace{-0.2cm}
  \hat{s}_i(x) = \frac{1}{1 + e^{\beta_i} \cdot e^{[\mathcal{F}(x)]_i}}
  \label{eqn:scaling_factor_function_of_ours}
\end{align}

\noindent
where $\beta_i$ represents one-dimensional learnable parameters along the output channel axis and $\mathcal{F(\cdot)}$ represents an arbitrary convolutional neural block.
Using Eq.~\ref{eqn:scaling_factor_function_of_ours}, all learnable parameters can have both negative and non-negative values, which have a higher degree of freedom than Eq.~\ref{eqn:normalized_scaling_factor_fucntion_of_gdn}. Moreover, input feature $x$ does not need to be squared, and the scaling factor function of Eq.~\ref{eqn:scaling_factor_function_of_ours} can rescale the different inputs that have the same magnitude but opposite sign with different scale values.

We directly compare the scaling factor function of Eq.~\ref{eqn:scaling_factor_function_of_GDN} from GDN and Eq.~\ref{eqn:scaling_factor_function_of_ours} of ours in Fig.~\ref{fig:GDN_limitation}. All points in Fig.~\ref{fig:GDN_limitation} are based on Joint Autoregressive~\cite{joint} models with only the normal version of GDN and the skip connection for stability. Red points represent Eq.~\ref{eqn:scaling_factor_function_of_GDN} from GDN, and blue points denote Eq.~\ref{eqn:scaling_factor_function_of_ours}.
The circle represents the models with only one $1\times1$ convolution, and the triangle is the model in which the $1\times1$ convolution is replaced by a $3\times3$ convolution. The rectangle represents the models with an additional $1\times1$ convolutional layers. We use the ReLU~\cite{relu} activation function between the $1\times1$ convolutional layers of Eq.~\ref{eqn:scaling_factor_function_of_GDN} from GDN to maintain the non-negative values, and we use a Leaky ReLU activation function for Eq.~\ref{eqn:scaling_factor_function_of_ours}. As we can see, in the case of Eq.~\ref{eqn:scaling_factor_function_of_GDN} from GDN, which is red points, even if the receptive field is expanded or more layers are added, the performance does not increase since the square root term limits the representation power or degree of freedom of the scaling module. In contrast, Eq.~\ref{eqn:scaling_factor_function_of_ours}, which is blue points, shows steady performance improvements as the scaling module expands. Additionally, the green circle point in Fig.~\ref{fig:GDN_limitation} represents Eq.~\ref{eqn:scaling_factor_function_of_ours} with a $3\times3$ convolution and the squared input. We can confirm that the squared input limits the representation power or degree of freedom of the network and decreases the performance. Therefore, we can confirm that replacing Eq.~\ref{eqn:scaling_factor_function_of_GDN} with Eq.~\ref{eqn:scaling_factor_function_of_ours} allows networks to overcome the limitations of representation power or degree of freedom.

\begin{figure}[t]
\centering
\begin{subfigure}{.94\textwidth}
    \centering
    \includegraphics[width=.99\linewidth]{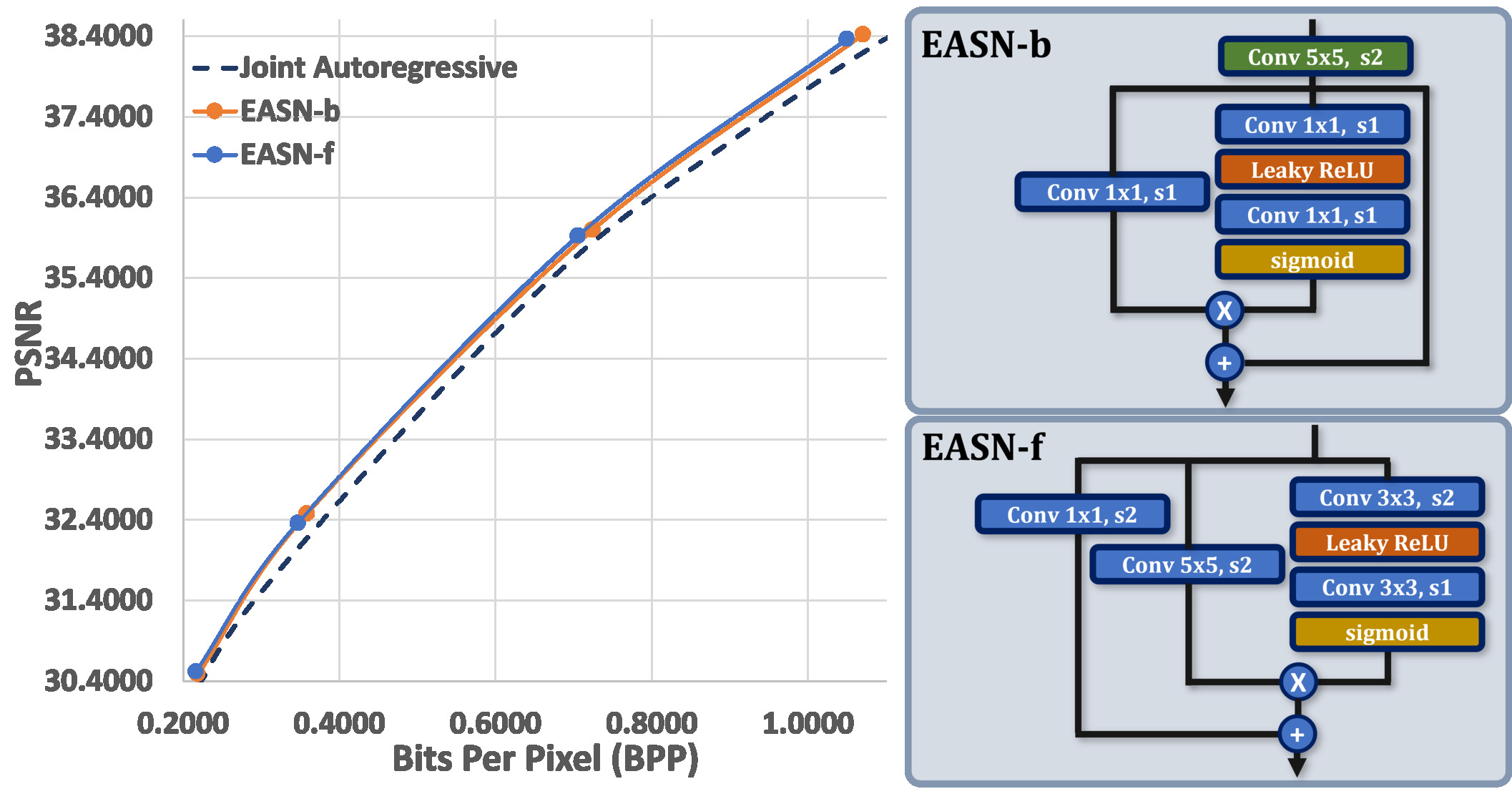}
\end{subfigure}
\caption{Comparison results with feature location. EASN-f uses features before down or upsampling for scaling factor function and input mapping function. Both models have a $5\times5$ receptive field, including down or upsampling layer.}
\label{fig:before_after}
\end{figure}

\subsection{EASN}
\vspace{-0.2cm}
GDN~\cite{end-to-end} has a single $1\times1$ convolution layer. Thus, GDN cannot deal with the spatial correlation, which is an important key in compression to reduce the spatial redundancy, and cannot approximate a more complex function. Since we can now utilize full representation power with a swish~\cite{swish} function from Sec.~\ref{sec:method-swish_function}, the scaling module can be expanded to consider the spatial correlation or obtain a higher degree of freedom.

We use two $3\times3$ convolutions with an intermediate Leaky ReLU activation function for the scaling factor function $\hat{s}(x)$ to increase the receptive field and make a function $\hat{s}(x)$ to be more complex. For these expansions, we add a skip connection to stabilize the training. Without a skip connection, such expansions make training unstable and training loss diverges very early. In many works~\cite{spatiotemporal,bam,cbam,res_ch_attn,second_order,holistic,attn_in_attn}, a skip connection is used to ensure stability when two different features are multiplied in the neural network. Therefore, we use a skip connection. Furthermore, we introduce another function, input mapping function $m(x)$, to provide the scaling module with the option of transforming the input features to increase the degree of freedom. We call this rescaling module, Expanded Adaptive Scaling Normalization(EASN), and the final equation for EASN is as follows.

\begin{align}
  EASN(x)=m(x) \cdot \hat{s}(x) + x
  \label{eqn:expanded_version}
\vspace{-0.4cm}
\end{align}

Furthermore, we find that it is useful to utilize features before down or upsampling for both the scaling factor function and input mapping function. If we use these features, we can get slightly better performance even with the same receptive field. If we compare two different models in Fig.~\ref{fig:before_after} that have the same $5 \times 5$ receptive field including the down or upsampling layer, the performance of EASN-f is shown to be slightly better than the EASN-b model.

Finally, we find that simply adding more layers to the scaling module does not efficiently lead to performance increases from ablation study results. Therefore, we propose a deeper EASN module called EASN-deep to obtain higher performance. As shown on the right-hand side of Fig.~\ref{fig:modules}, we cascade the EASN-f(front) and EASN-e(back) modules from Fig.~\ref{fig:ablation} that have a $5\times5$ convolution for the input mapping function. With this scheme, EASN-deep rescales the input feature twice, which leads to performance increases more efficiently. More experimental details of EASN-deep are demonstrated in Sec.~\ref{sec:ablation_study}.

\section{Experiments}
\vspace{-0.2cm}
\subsection{Implementation Details}
\vspace{-0.2cm}
We use MSE loss or MS-SSIM~\cite{msssim} loss to measure distortion for each PSNR or MS-SSIM performance comparison. Total loss is given as follows.

\begin{align}
    L_{total} = -\mathbb{E}[log_2 P] + \lambda \cdot D(x, \hat{x})
\end{align}

\noindent
where $P$ is estimated PMF, $x$ is the original image, and $\hat{x}$ is the reconstructed image.
In case of MSE loss for $D$, we use $D(x, \hat{x}) = 255^2 \cdot MSE(x, \hat{x})$, and for MS-SSIM loss, we use $D(x, \hat{x}) = (1 - MSSSIM(x, \hat{x}))$.
We set Hyperprior~\cite{hyperprior} and Joint Autoregressive~\cite{joint} model as the baseline. In case of EASN, we replace normal and inverse GDN~\cite{end-to-end} of baseline with ours. For EASN-deep, we replace both down or upsampling convolution and GDN with ours, because the down or upsampling process is included in the EASN-deep module.
Rate distortion trade-off parameter $\lambda$ is set to [0.005, 0.010, 0.020, 0.035, 0.080, 0.180] for MSE distortion loss, and to [7, 15, 30, 48, 110, 220] for MS-SSIM distortion loss.
$N$ is the base channel number, and $M$ is the output channel number of the latent representation $y$. For the Hyperprior baseline, we select $N=128$, $M=192$ for the front two $\lambda$ values. For the other $\lambda$ values, we set $N=192$, and $M=320$. For the Joint Autoregressive baseline, we set $N=192$, $M=192$ for the front two $\lambda$ values and we select $N=192$, $M=320$ for the other $\lambda$ values.

\subsection{Training}
\vspace{-3mm}
Basically, we follow the training process of CompressAI~\cite{compressai} framework. We use Vimeo90K~\cite{vimeo} dataset for training. We randomly crop training images into $256\times256$ size, and randomly flip them horizontally. We use Adam optimizer~\cite{adam} with batch size of 16, and learning rate is set to $1e^{-4}$ initially. We evaluate every epoch using validation set of COCO~\cite{coco} dataset to get the total loss of validation dataset. We crop the COCO validation dataset at the center with a size of $256\times256$. We reduce learning rate of factor $0.5$ if the validation loss does not improve during 10 epochs. We stop training when the learning rate decrease 4 times. In case of MS-SSIM loss, we fine-tune the pretrained model with MSE loss using initial learning rate of $0.5e^{-4}$ and stop training when the learning rate decreases 3 times.

\subsection{Evaluation}
\vspace{-3mm}
We use Kodak dataset~\cite{kodak} for evaluation for both PSNR and MS-SSIM metrics. Moreover, we use CLIC2021 validation dataset~\cite{clic}, which consists of 41 high resolution images for confirming robustness for more high resolution images. To evaluate rate-distortion performance, we measure bits per pixel (bpp). We save the bitstreams to a hard disk drive to get a physical file size and divide the size with the total pixels number of the image to get bpp. We draw rate-distortion (RD) curves to check the compression performance.

\begin{figure}[t!]
\centering
\begin{minipage}{0.99\linewidth}
\centering

\begin{subfigure}{.50\textwidth}
    \centering
    \includegraphics[width=.99\linewidth]{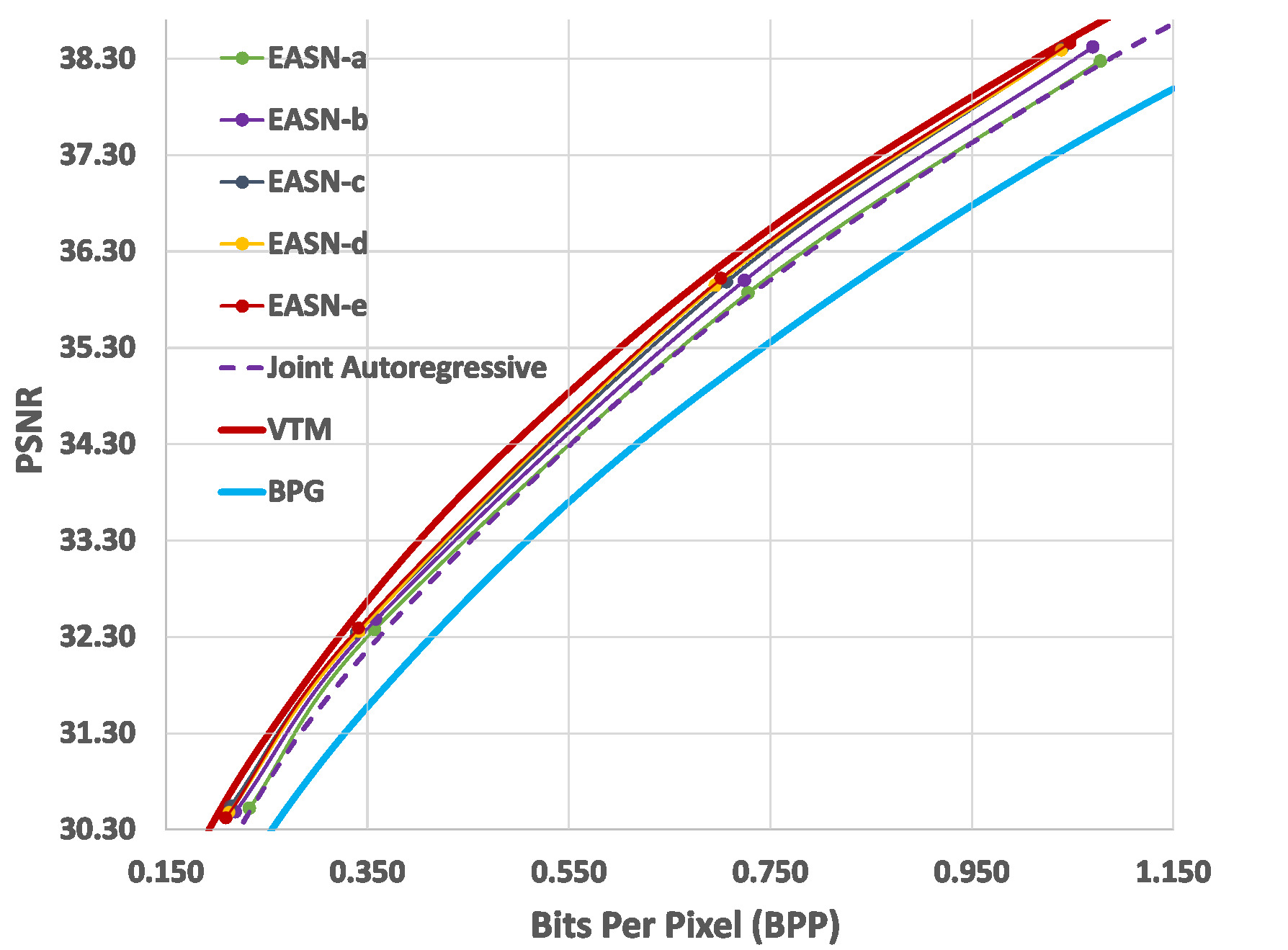}
    \caption{Ablation of EASN}
    \label{fig:ablation of EASN}
\end{subfigure}\hfill
\begin{subfigure}{.48\textwidth}
    \centering
    \includegraphics[width=.99\linewidth]{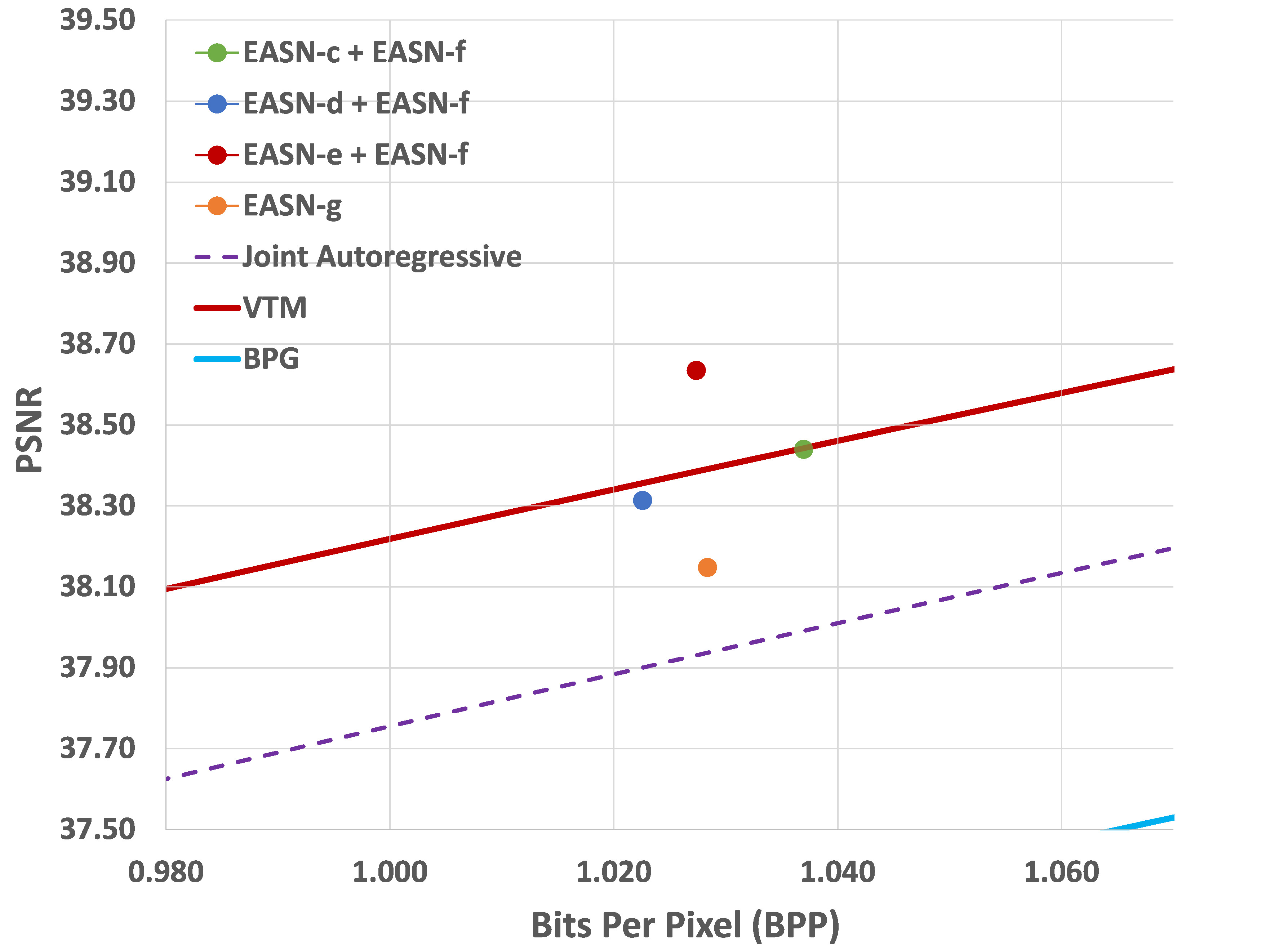}
    \caption{Ablation of EASN-deep}
    \label{fig:ablation of EASN-deep}
\end{subfigure}

\end{minipage}
\vspace{4mm}\vfill
\begin{minipage}{0.99\linewidth}
    \centering
    \resizebox{0.85\linewidth}{!}
{
    \begin{tabular}{c||c|c|c|c|c}
              & \textbf{EASN-a} & \textbf{EASN-b} & \textbf{EASN-c} & \textbf{EASN-d} & \textbf{EASN-e}  \\ \hline\hline
$\hat{s}(x)$ & {[} 1 $\times$ 1 {]} $\times$ 2  & {[} 1 $\times$ 1 {]} $\times$ 2  & {[} 3 $\times$ 3 {]} $\times$ 2  & {[} 3 $\times$ 3 {]} $\times$ 2  & {[} 3 $\times$ 3 {]} $\times$ 2  \\ \hline
$m(x)$ & $I$              & {[} 1 $\times$ 1 {]} $\times$ 1  & {[} 1 $\times$ 1 {]} $\times$ 1  & {[} 1 $\times$ 1 {]} $\times$ 1  & {[} 5 $\times$ 5 {]} $\times$ 1 \\ \hline
$h(x)$ & 0         & 0                         & 0             & {[} 1 $\times$ 1 {]} $\times$ 1           & 0            
\end{tabular}
}
\end{minipage}
\caption{Ablation study result. [ k $\times$ k ] $\times$ L means using L number of k $\times$ k convolution. $I$ is identity function. 0 means multiplying zero.}
\label{fig:ablation}
\end{figure}

\subsection{Ablation Study}\label{sec:ablation_study}
Fig.~\ref{fig:ablation of EASN} represents the ablation study results of EASN, and Fig.~\ref{fig:ablation of EASN-deep} shows the results of EASN-deep. We expand existing GDN~\cite{end-to-end} to the following equation.

\begin{align}
  EASN(x) = m(x) \cdot \hat{s}(x) + h(x) + x
\end{align}

\noindent where $h(x)$ is shift function. The table from Fig.~\ref{fig:ablation} shows module structure of each modules. $[k\times k] \times L$ represents that it has L number of $k \times k$ convolutions. $I$ is identity function, and $0$ means that those modules do not use a corresponding function.
GDN~\cite{end-to-end} and EASN modules with skip connection have a worse performance than original GDN of Joint Autoregressive~\cite{joint}. However, if we look at the Fig.~\ref{fig:ablation of EASN}, we can make up slightly poor performance using only one more $1\times1$ convolution (EASN-a). EASN-a from Fig.~\ref{fig:ablation} shows the same performance as the GDN-based Joint Autoregressive~\cite{joint} model. As we add $1\times1$ convolution to input mapping function $m(x)$ (EASN-b), and replace $1\times1$ convolution of scaling factor function $\hat{s}(x)$ with $3\times3$ convolution (EASN-c), performance of the EASN modules increase steadily. In case of EASN-d, we use shift function with $1 \times 1$ convolution. However, we find that the performance does not increase. EASN-e has $5 \times 5$ convolution for input mapping function. Although the performance slightly increases, but considering the parameter numbers, we select EASN-c for the final EASN.

Fig.~\ref{fig:ablation of EASN-deep} shows the performance comparison results of combining EASN-f with EASN-c, EASN-d, and EASN-e, which show the highest performance within EASN ablation results. In case of EASN-g, we simply add two more $3 \times 3$ convolution layers to scaling factor function $\hat{s}(x)$ of EASN-f and one more $5 \times 5$ convolution layer to input mapping function $m(x)$ of EASN-f to make same receptive field as the EASN-e + EASN-f module. As shown in Fig.~\ref{fig:ablation of EASN-deep}, we can confirm that simply adding more layers decreases the performance. Therefore, for constructing a deeper adaptive rescaling module effectively, we cascade the EASN-f module with other EASN modules. We find that the receptive field of input mapping function is important in terms of cascading two modules. Using $5 \times 5$ convolution for input mapping function (EASN-e) shows significant performance improvement. Therefore, we choose the combination of EASN-f(front) and EASN-e(back) modules for EASN-deep version.

\begin{figure}[t]
\centering
\begin{subfigure}{.48\textwidth}
    \centering
    \includegraphics[width=.99\linewidth]{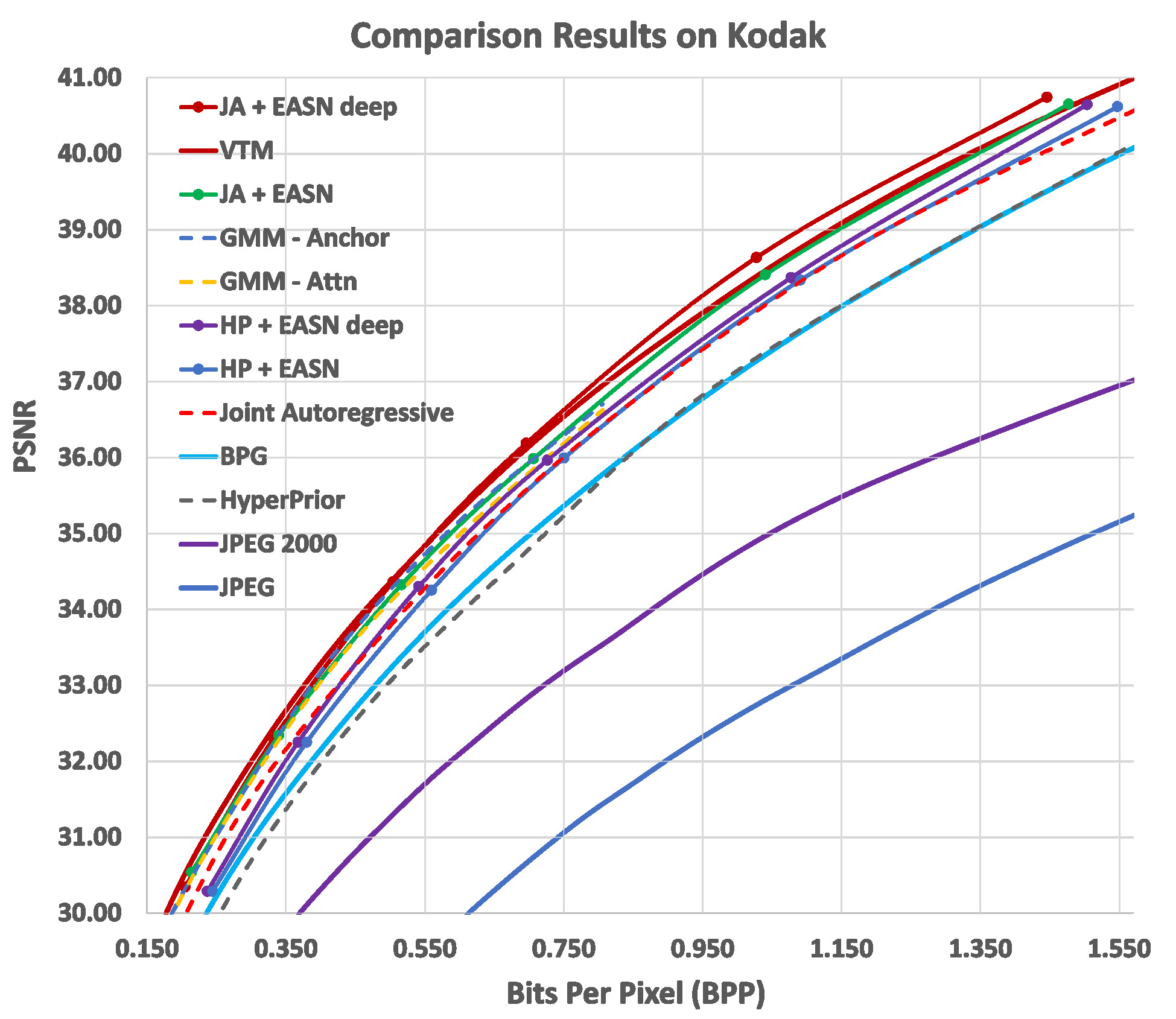}
\end{subfigure}\hfill
\vspace{3mm}
\begin{subfigure}{.48\textwidth}
    \centering
    \includegraphics[width=.99\linewidth]{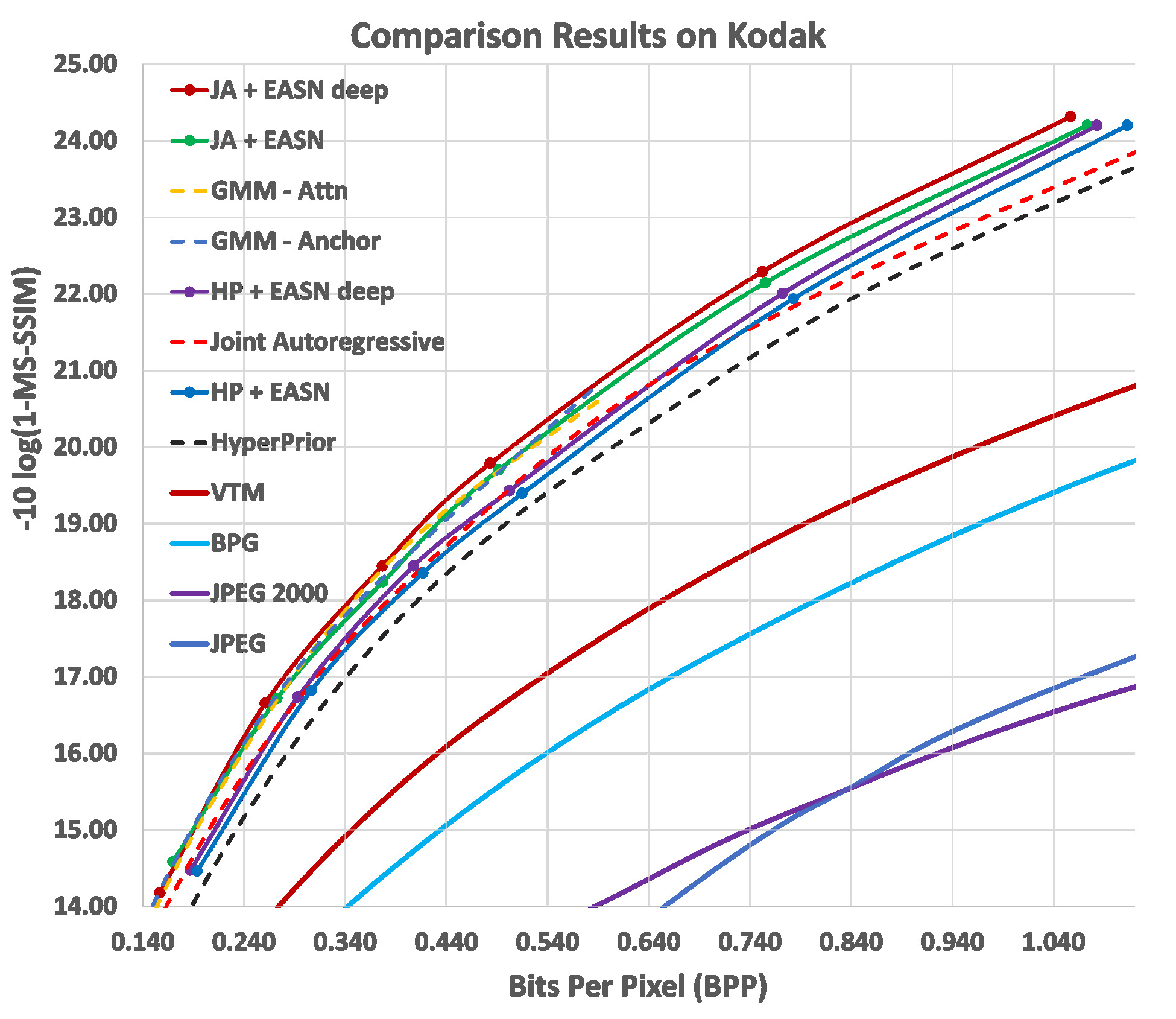}
\end{subfigure}\vfill

\begin{subfigure}{.48\textwidth}
    \centering
    \includegraphics[width=.99\linewidth]{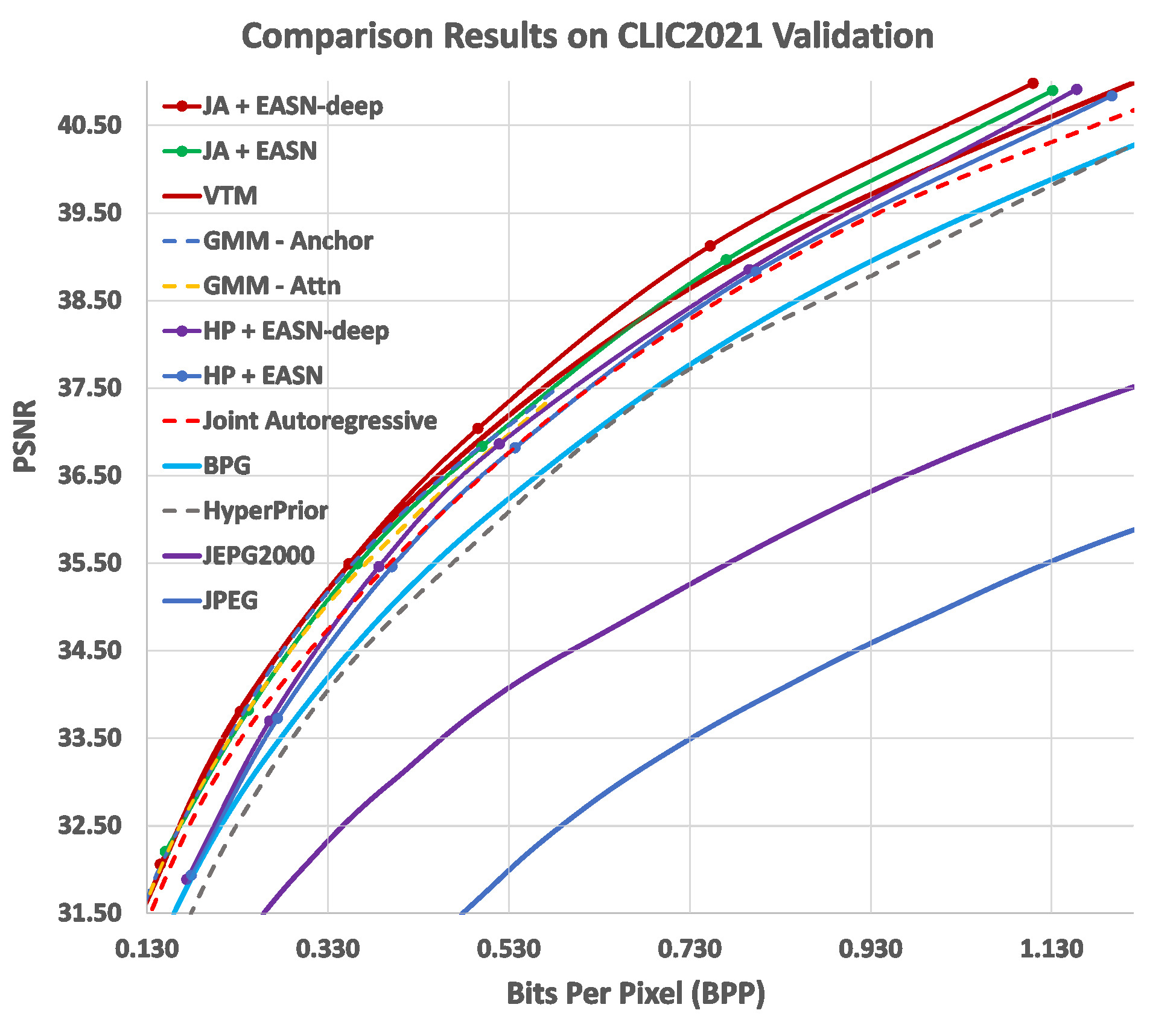}
    \caption{RD performance with PSNR}
    \label{fig:psnr result}
\end{subfigure}\hfill
\vspace{3mm}
\begin{subfigure}{.48\textwidth}
    \centering
    \includegraphics[width=.99\linewidth]{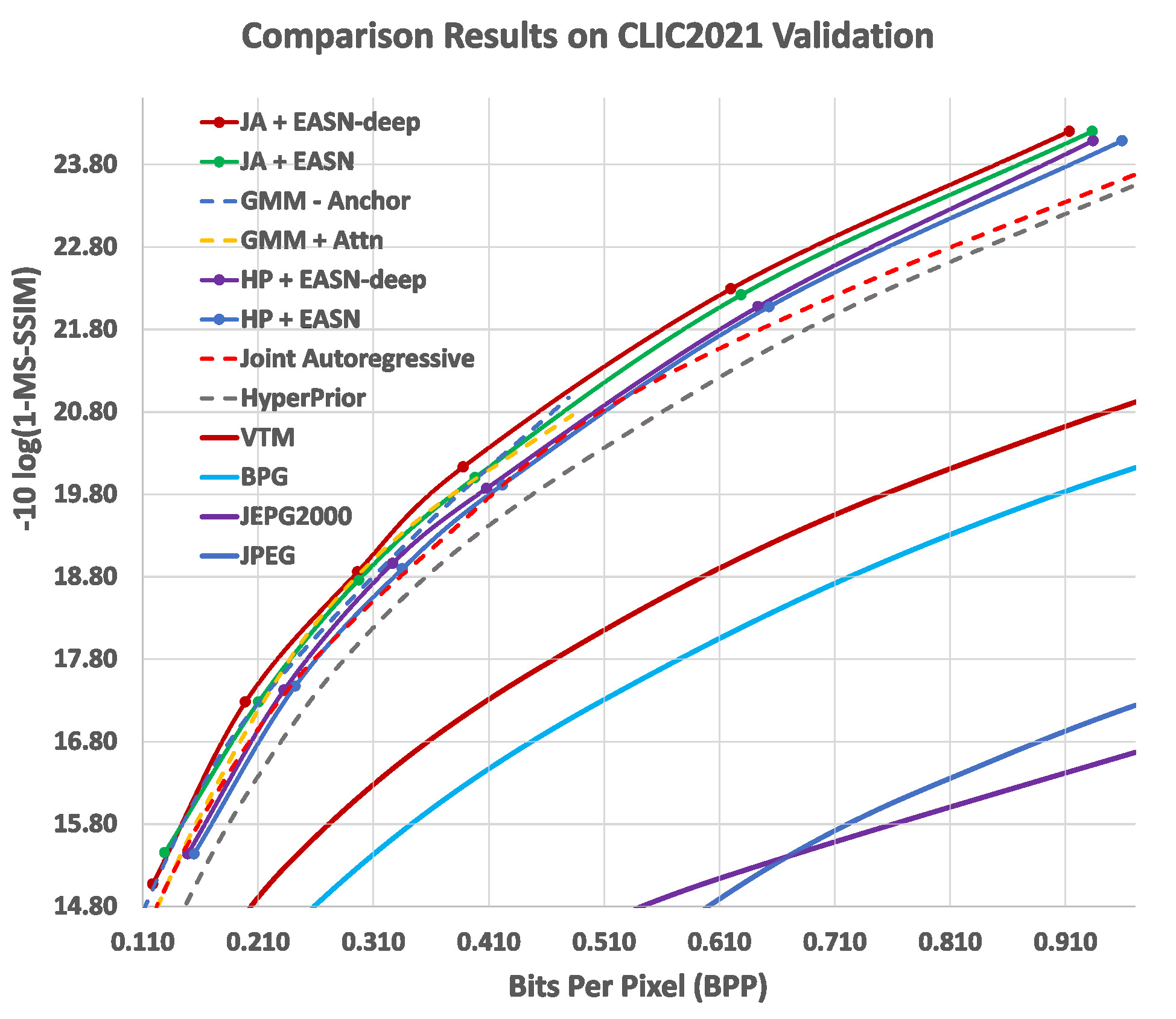}
    \caption{RD performance with MS-SSIM}
    \label{fig:msssim result}
\end{subfigure}
\vspace{-2mm}
\caption{Rate-distortion Performance comparison results on Kodak~\cite{kodak} and CLIC2021 validation dataset~\cite{clic}.}
\label{fig:kodak result}
\vspace{-0.5cm}
\end{figure}

\subsection{Rate Distortion Performance}
\vspace{-2mm}
For comparison, we use traditional codecs of JPEG~\cite{JPEG}, JPEG2000~\cite{JPEG2000}, BPG~\cite{BPG} which is image codec based on HEVC~\cite{hevc}, and VTM~\cite{VTM} which is the official test model of VVC~\cite{vvc}. For learned image compression method, we use HyperPrior~\cite{hyperprior}, Joint Autoregressive~\cite{joint} and GMM~\cite{gmm}. For GMM, we use two different version, Anchor and Attention. The only difference between them is existence of the attention module. We adapt our EASN and EASN-deep to HyperPrior and Joint Autoregressive models which have a GDN-based structure. We plot two separate figures optimized by MSE or MS-SSIM~\cite{msssim}, respectively. In case of MS-SSIM, we use log scale for visualization.

Fig.~\ref{fig:psnr result} shows the rate-distortion performance with PSNR metric on both dataset. HP and JA represent HyperPrior and Joint Autoregressive, respectively. As we can see, HP + EASN outperforms HyperPrior that has a similar performance with traditional BPG, and HP + EASN-deep model shows higher performance than the HP + EASN model. The JA + EASN model outperforms the Joint Autoregressive model, and even shows similar performance with GMM Anchor model. At high bit rate, Our JA + EASN model reach the rate-distortion performance of traditional codec of VTM. In case of JA + EASN-deep model, although its performance is similar with JA + EASN model at low bit rate, our model outperforms all other learning-based and traditional codecs at high bit rate on both datasets. Fig.~\ref{fig:msssim result} shows the performance comparison results with MS-SSIM metric. They show a similar tendency to PSNR results. HP + EASN and JA + EASN models outperform the baselines of HyperPrior and Joint Autoregressive, respectively. The HP + EASN-deep model has a higher performance than our HP + EASN models, and our JA + EASN-deep model outperforms all other learning-based models on both datasets.

\begin{figure}[t!]
\centering
\begin{subfigure}{.98\textwidth}
    \centering
    \includegraphics[width=.99\linewidth]{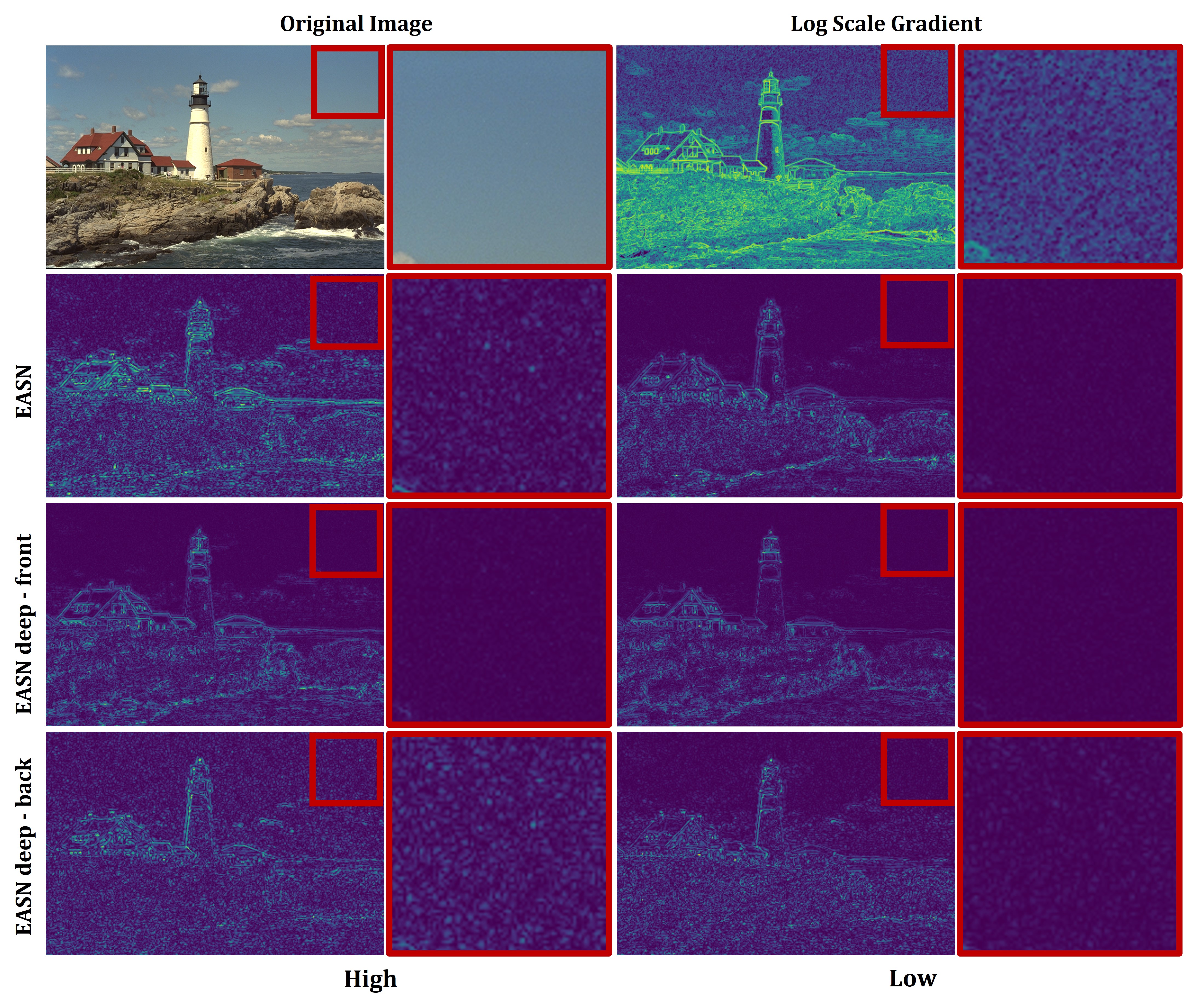}
\end{subfigure}
\caption{Visualization of high frequency components of scaling factor function output $\hat{s}(x)$ with kodim21 image from Kodak dataset~\cite{kodak}. The top-right image represents the log scale gradient result of original image. The left column images denotes high bit rate results, and right column images represents low bit rate results.}
\label{fig:feature map visualization}
\end{figure}

\subsection{Scale Feature Map}
\vspace{-2mm}
In this section, we demonstrate how the EASN module works along the bit rates with the visualization results of feature maps of the scaling factor function $\hat{s}(x)$. A low bit rate model discards many high frequency information to obtain a high compression rate. Whereas, a high bit rate model should generate reconstructed images with low distortion that comprise many fine details. To confirm the difference between various bit rate models, we remove low-frequency components to focus on high frequency details. Intuitively, the two models with different bit rates may rescale the blue color pixels with different values, such as 0.1 and 0.8, respectively. Therefore, the exact scale values are not important, and we should focus on the variety of scale values in accordance with the pixel variety of input images. We remove the low-frequency components using the following equation.

\vspace{-0.5cm}
\begin{align}
    x_{avg}^{hf}=\frac{1}{N}\sum_{n=0}^Nx_n - (x_n*k_{3\times3})
\end{align}
\vspace{-0.2cm}

\noindent
where $k_{3\times3}$ is the mean filter with a kernel size of $3\times3$, and $1/9$ value for all components. Symbol $*$ is a convolution operator, $n$ indicates channel axis, and $N$ is the channel number. $x$ is the feature map from the scaling factor function $\hat{s}(x)$ of the first EASN module in the encoder.

Fig.~\ref{fig:feature map visualization} is the visualization results of the high frequency components of the feature $x$ of the scaling factor function $\hat{s}(x)$ with the kodim21 image from the Kodak dataset~\cite{kodak}. The top-left image is the original image, and the top-right image is the gradient of the original image with log scale. The vertical axis represents each module, the left column images represent high bit rate models, and the right column images represent low bit rate models. For EASN-deep model, there are two scaling factor functions. EASN-deep front is the first rescaling part, which is the EASN-f module, and EASN-deep back is the second rescaling part, which is the EASN-e module represented in Fig.~\ref{fig:modules}.

The log scale gradient of the original image shows high frequency components in the sky of the image, which is a flat region. Unlike the textures or edges, these details are not clearly visible to the human eye. The high bit rate module results of the EASN show that they catch these details in the red boxes of the sky region. In the case of GDN-deep front, this does not show a difference for the sky region but the EASN-deep back shows high frequency components in the sky. In contrast, there are no high frequency components in the sky region for all models trained for low bit rate. This means that models trained with a high bit rate catch more fine details in images. 
From these results, we can confirm that the scaling factor function $\hat{s}(x)$ in our EASN can adjust and rescale high frequency components of input features depending on the bit rates. We can also interpret these results as the scaling factor function $\hat{s}(x)$ determines how many details to remove to save bits.

\section{Conclusions}
We propose Expanded Adaptive Scaling Normalization(EASN), which is an expanded structure of existing GDN. For constructing EASN, first we exploit the swish function for the scaling factor function to make the module to utilize representation power fully. Second, we increase receptive field and make the scaling factor function deeper to consider spatial correlation and approximate more complex function. Additionally, we add input mapping function to increase degree of freedom, and we propose more EASN-deep module to make the module more deeper effectively. Furthermore, we reveal the process of how our EASN works along the bit rates within an image compression network using the visualization results of feature map. We conduct extensive experiments to show that each of the proposed methods is effective through ablation study, and our EASN shows dramatic increase of performance, and even outperforms other image compression methods.

\vspace{8mm}
\noindent
\scriptsize{\textbf{Acknowledgement}. This work was supported by Institute of Information \& communications Technology Planning \& Evaluation (IITP) grant funded by the Korea government(MSIT) (No.2021-0-02068, Artificial Intelligence Innovation Hub)}

\clearpage
\bibliographystyle{splncs04}
\bibliography{egbib}

\pagebreak
%\begin{comment}
\title{Supplementary Material} % Replace with your title
\titlerunning{EASN Supplementary Material}
\author{Chajin Shin\index{Shin, Chajin} \and
Hyeongmin Lee \and
Hanbin Son \and
Sangjin Lee \and
Dogyoon Lee\and
Sangyoun Lee
}
\authorrunning{C. Shin et al.}
\institute{School of Electrical and Electronic Engineering, Yonsei University, Seoul, Korea}
%\end{comment}
%******************
\maketitle

\section{Traditional Codec Settings}
In this section, we describe test settings of traditional codecs of JPEG~\cite{JPEG}, JPEG2000~\cite{JPEG2000}, BPG~\cite{BPG}, and VTM~\cite{VTM}. In case of JPEG, we use Pillow~\cite{pillow}, which is an imaging library of Python~\cite{python} programming language, to control bit rate of JPEG images.
For JPEG2000, we utilize FFmpeg~\cite{ffmpeg} 3.5.8 version which is open-source library that handle video, image, and audio. We use following command line to encode images.

\vspace{0.5cm}
\noindent
\texttt{ffmpeg -i input.jpg -vcodec jpeg2000 -pix\_fmt yuv444p -c:v\\libopenjpeg -compression\_level QP output.jp2}
\vspace{0.5cm}

\noindent
where \texttt{input.jpg} is the original input image directory, \texttt{QP} is quality factor, and we set QP as [5, 10, 15, 25, 35, 45, 55, 65, 75, 85]. The \texttt{output.jp2} is compressed result directory. To decode compressed image, we use following command line.

\vspace{0.5cm}
\noindent
\texttt{ffmpeg -i input.jp2 output.jpg}
\vspace{0.5cm}

In case of BPG, we use libbpg~\cite{libbpg} 0.9.5 version and following command line to encode original image.

\vspace{0.5cm}
\noindent
\texttt{bpgenc -o output.bpg -q QP -f 444  -e x265 -c ycbcr -b 8 input.jpg}
\vspace{0.5cm}

\noindent
We set QP as [15, 20, 25, 30, 35, 40, 45]. For decoding compressed result of \texttt{output.bpg}, we use following command line.

\vspace{0.5cm}
\noindent
\texttt{bpgdec -o output.jpg input.bpg}
\vspace{0.5cm}

In case of VTM 9.1 version which is the official test module of VVC~\cite{vvc}, we first convert RGB space to YUV444 space, and we encode input image with following command line.

\vspace{0.5cm}
\noindent
\texttt{EncoderApp -i input.yuv -c encoder\_intra\_vtm.cfg -q QP -o output.yuv -b output.bin -wdt width -hgt height -fr 1 -f 1 --InputChromaFormat\\=444 --InputBitDepth=8}
\vspace{0.5cm}

\noindent
where \texttt{encoder\_intra\_vtm.cfg} is default intra configuration file, and we set QP as [15, 20, 25, 30, 35, 40, 45]. The notation \texttt{width} and \texttt{height} are width and height of original input image, respectively, and \texttt{output.bin} is compressed results. We do not use \texttt{output.yuv} from encoder result, and we decode compressed result of \texttt{output.bin} to reconstruct the image with following command line.

\vspace{0.5cm}
\noindent
\texttt{DecoderApp -b input.bin -o output.yuv -d 8}
\vspace{0.5cm}

\noindent
After decoding compressed file to get reconstructed YUV file, we convert YUV444 space to RGB space to measure distance with the original image.

\newpage
\section{Scale Feature Map}
In this section, we visualize the output feature of the scaling factor function $\hat{s}(x)$ with kodim07 and kodim09 images from Kodak datset~\cite{kodak}. In case of kodim07 image, the top-left image represents the original image and the top-right image represents the log scale gradient result of the original image. The left column images represent high bit rate results, and the right column images represents low bit rate results. In case of kodim09 image, the top-left image represents the original image and bottom-left image is the log scale gradient result of the original image. The top row indicate the high bit rate models and bottom row represents the low bit rate models.

\begin{figure}[h!]
\centering
\begin{subfigure}{.98\textwidth}
    \centering
    \includegraphics[width=.99\linewidth]{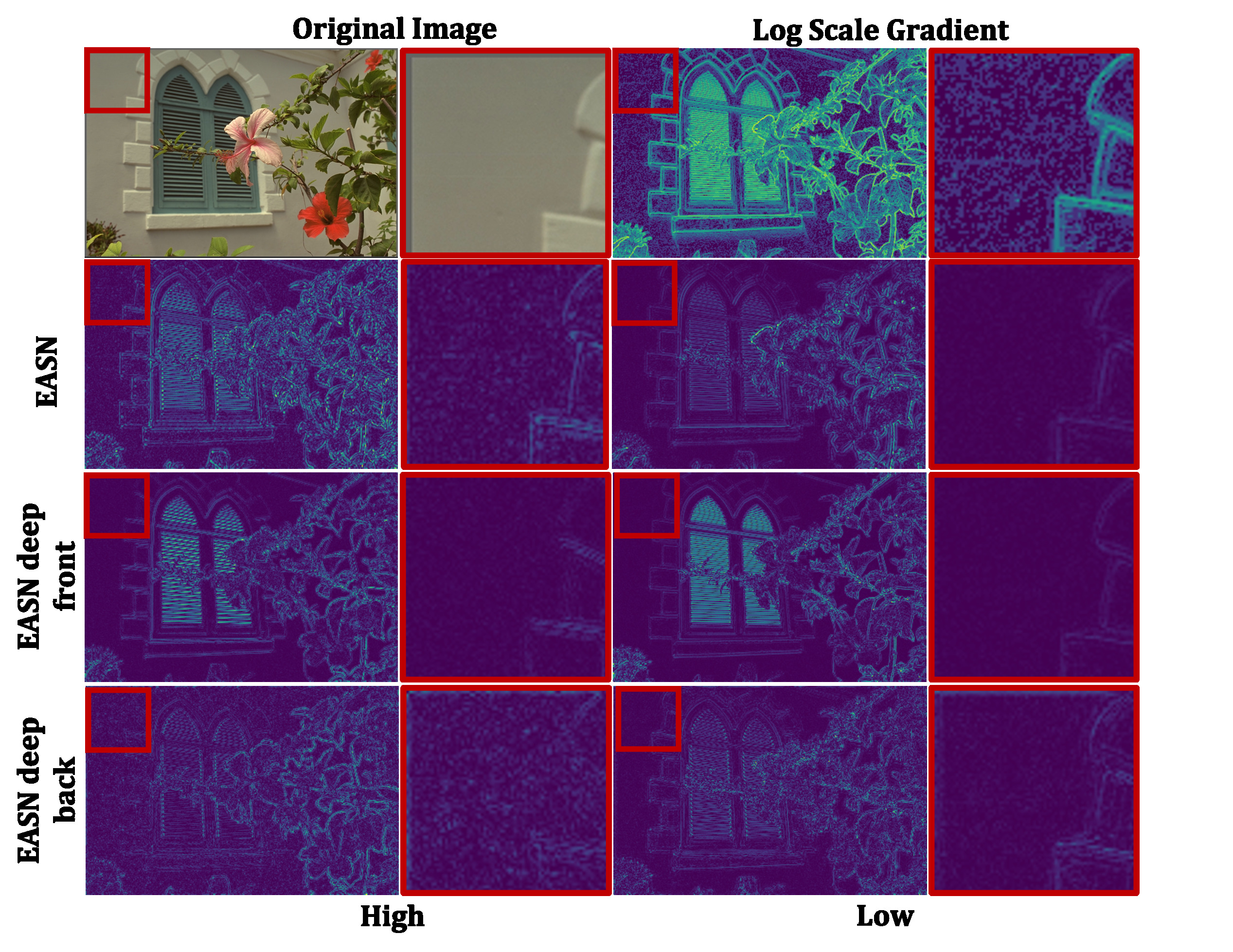}
\end{subfigure}
\caption{Visualization result with kodim07 image from Kodak dataset.}
\label{fig:feature map visualization kodim07}
\end{figure}

\begin{figure}[h!]
\centering
\begin{subfigure}{.98\textwidth}
    \centering
    \includegraphics[width=.99\linewidth]{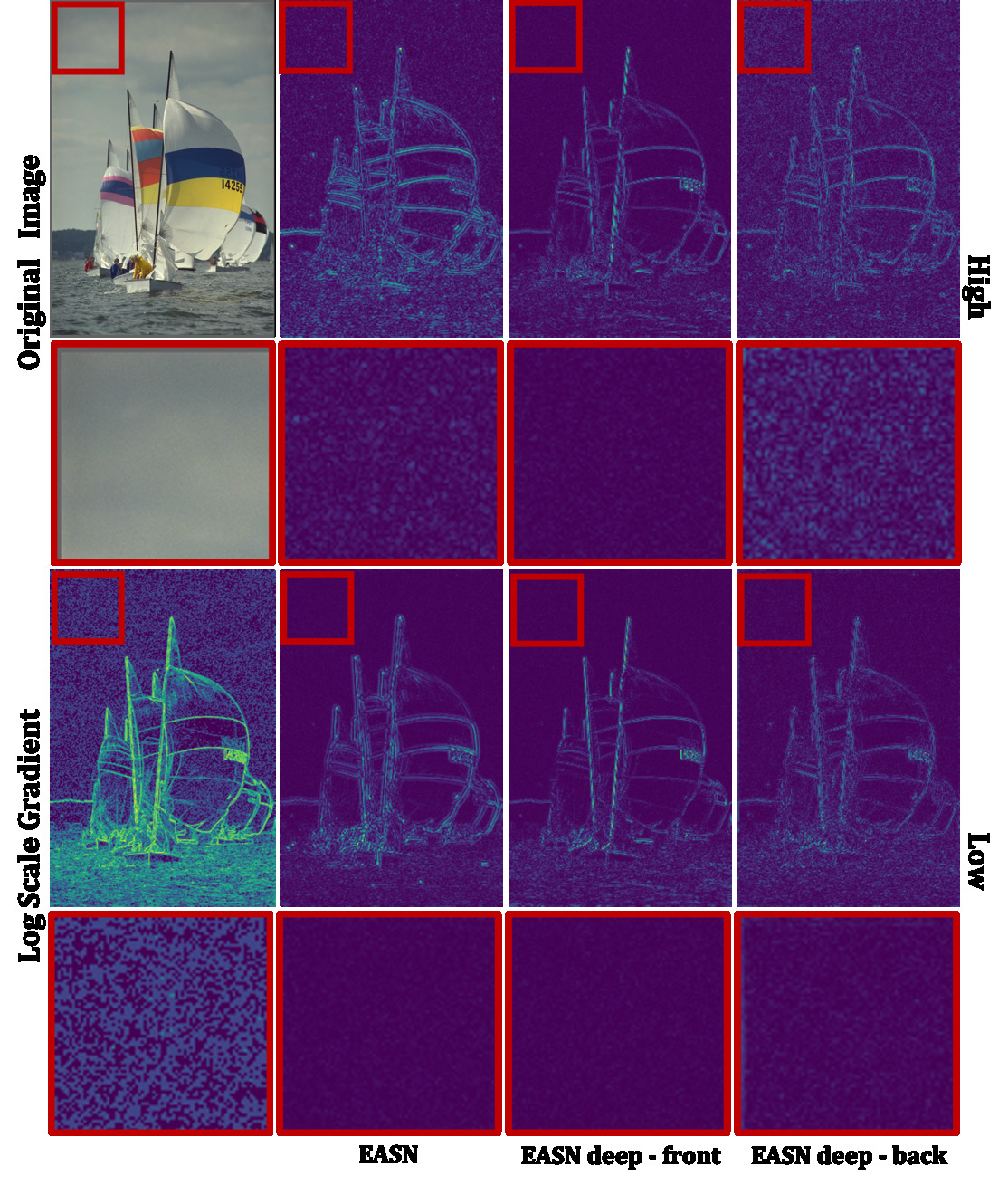}
\end{subfigure}
\caption{Visualization result with kodim09 image from Kodak dataset.}
\label{fig:feature map visualization kodim09}
\vspace{5cm}
\end{figure}

\clearpage
\section{Qualitative Results}
In this section, we qualitatively compare our best model of JA + EASN-deep with other traditional codecs. We can find that our model catch fine detail much better than other methods.

\begin{figure}[h!]
\centering
\begin{subfigure}{.98\textwidth}
    \centering
    \includegraphics[width=.99\linewidth]{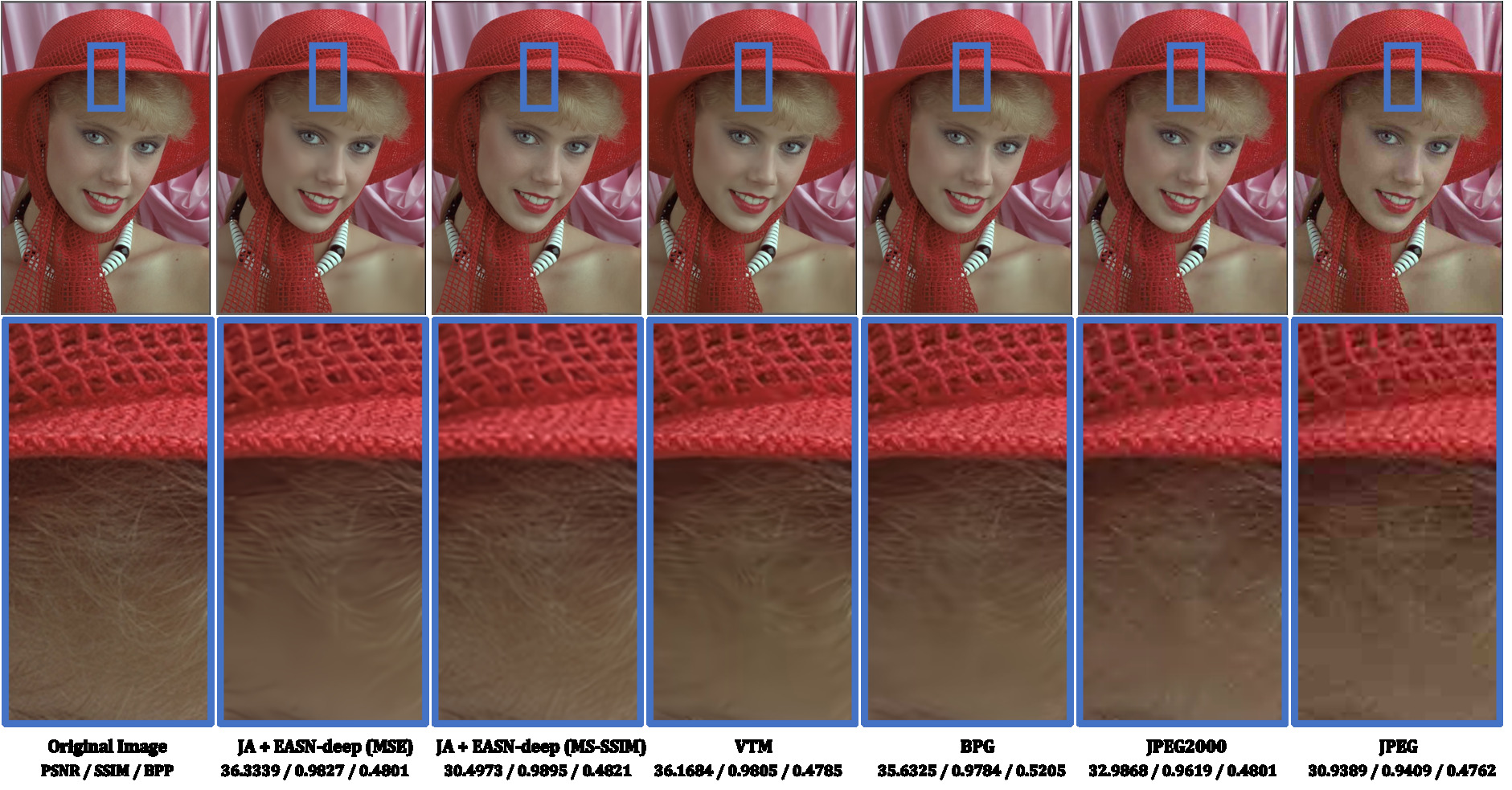}
\end{subfigure}
\caption{Qualitative comparison result with kodim04 image from Kodak dataset.}
\label{fig:qualitative comparison kodim04}
\end{figure}

\begin{figure}[h!]
\centering
\begin{subfigure}{.98\textwidth}
    \centering
    \includegraphics[width=.99\linewidth]{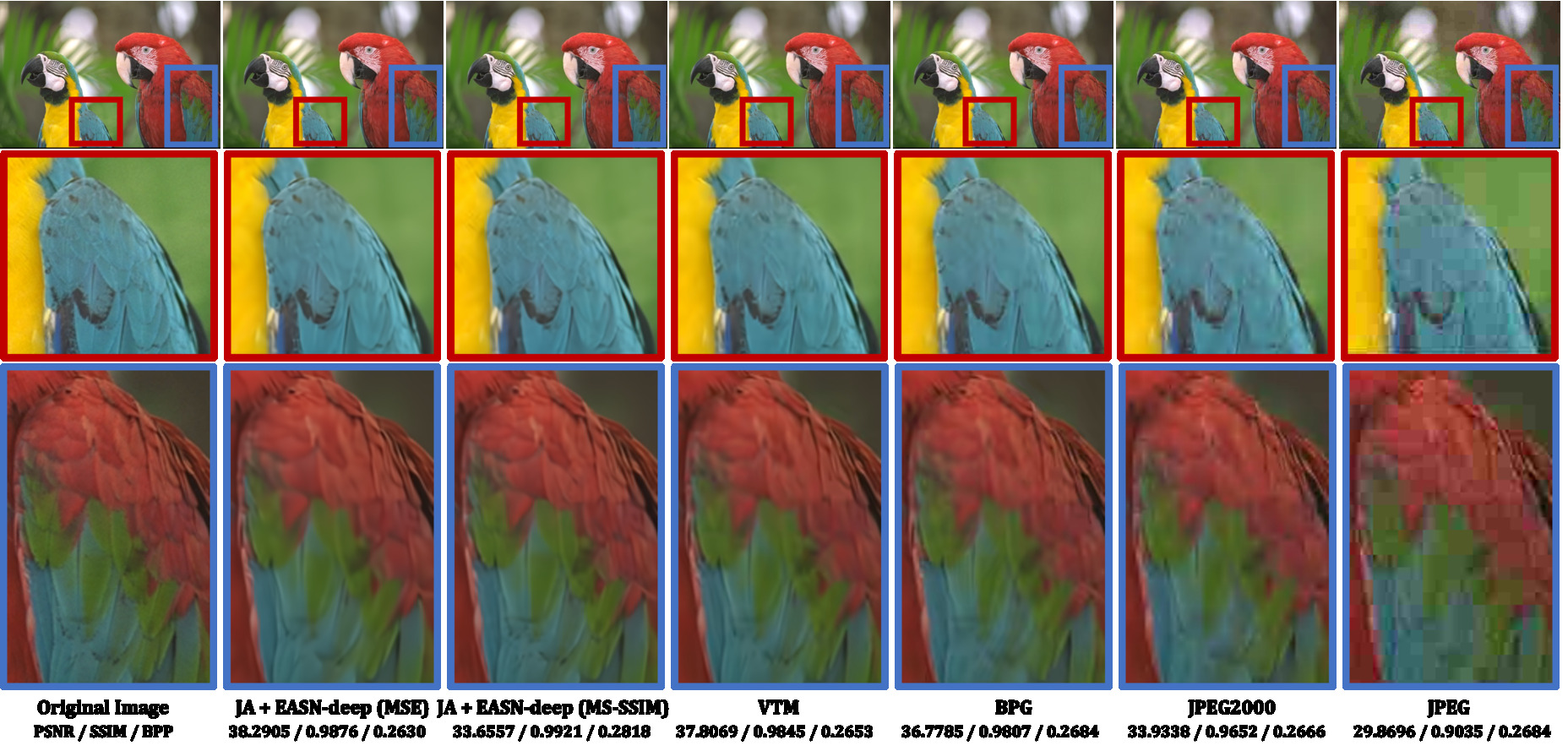}
\end{subfigure}
\caption{Qualitative comparison result with kodim23 image from Kodak dataset.}
\label{fig:qualitative comparison kodim23}
\end{figure}

\begin{figure}[t]
\begin{subfigure}{.98\textwidth}
    \centering
    \includegraphics[width=.99\linewidth]{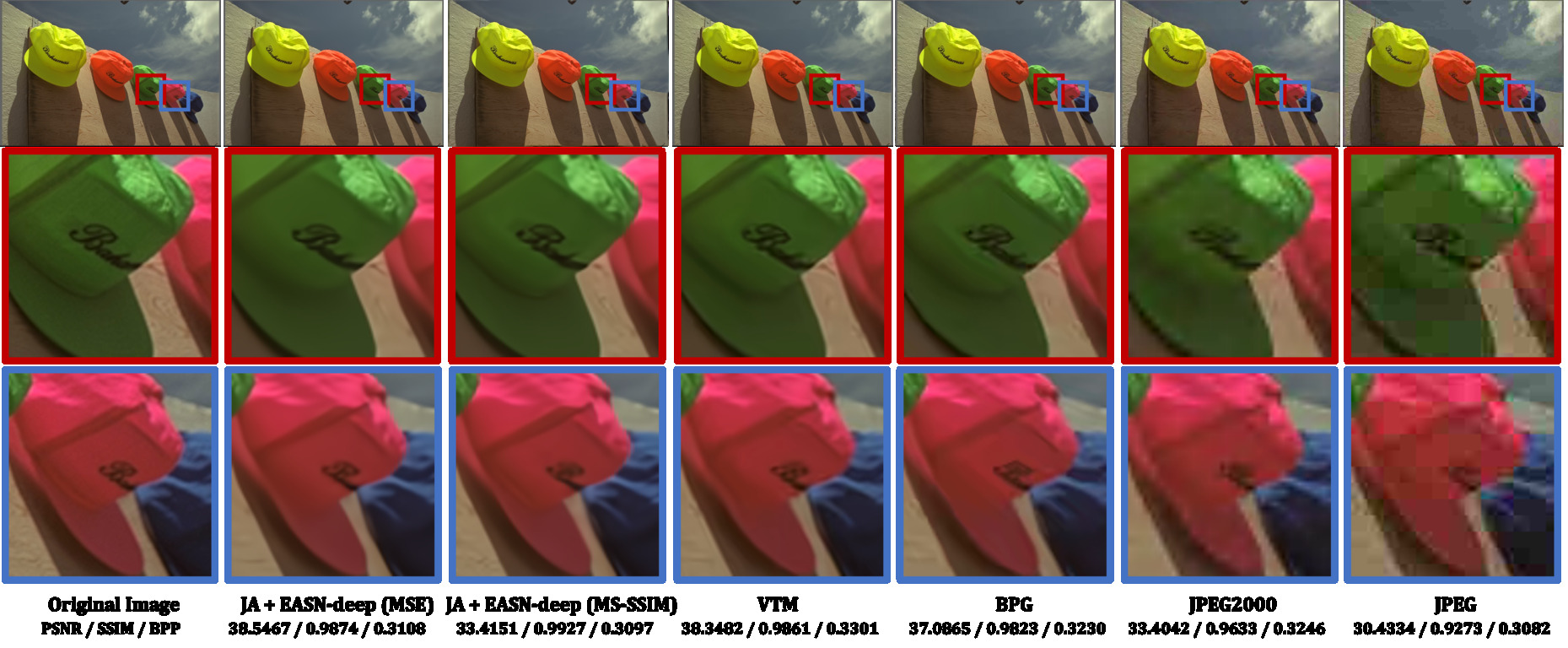}
\end{subfigure}
\caption{Qualitative comparison result with kodim03 image from Kodak dataset.}
\label{fig:qualitative comparison kodim03}
\vspace{13cm}
\end{figure}

%\clearpage\mbox{}Page \thepage\ of the manuscript.
%\clearpage\mbox{}Page \thepage\ of the manuscript.

%This is the last page of the manuscript.
%\par\vfill\par
%Now we have reached the maximum size of the ECCV 2022 submission (excluding references).
%References should start immediately after the main text, but can continue on p.15 if needed.

\clearpage
\bibliographystyle{splncs04}
\bibliography{egbib}

\begin{thebibliography}{10}
\providecommand{\url}[1]{\texttt{#1}}
\providecommand{\urlprefix}{URL }
\providecommand{\doi}[1]{https://doi.org/#1}

\bibitem{VTM}
\uppercase{VVC VTM} reference software
  \url{https://vcgit.hhi.fraunhofer.de/jvet/VVCSoftware\\\_VTM}

\bibitem{relu}
Agarap, A.F.: Deep learning using rectified linear units (relu). arXiv preprint
  arXiv:1803.08375  (2018)

\bibitem{end-to-end}
Ball{\'e}, J., Laparra, V., Simoncelli, E.P.: End-to-end optimized image
  compression. arXiv preprint arXiv:1611.01704  (2016)

\bibitem{hyperprior}
Ball{\'e}, J., Minnen, D., Singh, S., Hwang, S.J., Johnston, N.: Variational
  image compression with a scale hyperprior. arXiv preprint arXiv:1802.01436
  (2018)

\bibitem{compressai}
B{\'e}gaint, J., Racap{\'e}, F., Feltman, S., Pushparaja, A.: Compressai: a
  pytorch library and evaluation platform for end-to-end compression research.
  arXiv preprint arXiv:2011.03029  (2020)

\bibitem{BPG}
Bellard, F.: Bpg image format  (2015), \url{Signal processing: Image
  communication}

\bibitem{attn_in_attn}
Chen, H., Gu, J., Zhang, Z.: Attention in attention network for image
  super-resolution. arXiv preprint arXiv:2104.09497  (2021)

\bibitem{gmm}
Cheng, Z., Sun, H., Takeuchi, M., Katto, J.: Learned image compression with
  discretized gaussian mixture likelihoods and attention modules. In:
  Proceedings of the IEEE/CVF Conference on Computer Vision and Pattern
  Recognition. pp. 7939--7948 (2020)

\bibitem{asymmetric}
Cui, Z., Wang, J., Gao, S., Guo, T., Feng, Y., Bai, B.: Asymmetric gained deep
  image compression with continuous rate adaptation. In: Proceedings of the
  IEEE/CVF Conference on Computer Vision and Pattern Recognition. pp.
  10532--10541 (2021)

\bibitem{clic}
CVPR2021: Workshop and challenge on learned image compression  (2021),
  \url{http://clic.compression.cc/2021/tasks/index.html}

\bibitem{second_order}
Dai, T., Cai, J., Zhang, Y., Xia, S.T., Zhang, L.: Second-order attention
  network for single image super-resolution. In: Proceedings of the IEEE/CVF
  conference on computer vision and pattern recognition. pp. 11065--11074
  (2019)

\bibitem{spatiotemporal}
Feichtenhofer, C., Pinz, A., Wildes, R.P.: Spatiotemporal multiplier networks
  for video action recognition. In: Proceedings of the IEEE conference on
  computer vision and pattern recognition. pp. 4768--4777 (2017)

\bibitem{checkerboard}
He, D., Zheng, Y., Sun, B., Wang, Y., Qin, H.: Checkerboard context model for
  efficient learned image compression. In: Proceedings of the IEEE/CVF
  Conference on Computer Vision and Pattern Recognition. pp. 14771--14780
  (2021)

\bibitem{batchnorm}
Ioffe, S., Szegedy, C.: Batch normalization: Accelerating deep network training
  by reducing internal covariate shift. In: International conference on machine
  learning. pp. 448--456. PMLR (2015)

\bibitem{adam}
Kingma, D.P., Ba, J.: Adam: A method for stochastic optimization. arXiv
  preprint arXiv:1412.6980  (2014)

\bibitem{kodak}
Kodak, E.: Kodak lossless true color image suite (photocd pcd0992)
  \url{http://r0k.us/graphics/kodak/}

\bibitem{context}
Lee, J., Cho, S., Beack, S.K.: Context-adaptive entropy model for end-to-end
  optimized image compression. arXiv preprint arXiv:1809.10452  (2018)

\bibitem{coco}
Lin, T.Y., Maire, M., Belongie, S., Hays, J., Perona, P., Ramanan, D.,
  Doll{\'a}r, P., Zitnick, C.L.: Microsoft coco: Common objects in context. In:
  European conference on computer vision. pp. 740--755. Springer (2014)

\bibitem{joint}
Minnen, D., Ball{\'e}, J., Toderici, G.D.: Joint autoregressive and
  hierarchical priors for learned image compression. Advances in neural
  information processing systems  \textbf{31} (2018)

\bibitem{channel-wise}
Minnen, D., Singh, S.: Channel-wise autoregressive entropy models for learned
  image compression. In: 2020 IEEE International Conference on Image Processing
  (ICIP). pp. 3339--3343. IEEE (2020)

\bibitem{holistic}
Niu, B., Wen, W., Ren, W., Zhang, X., Yang, L., Wang, S., Zhang, K., Cao, X.,
  Shen, H.: Single image super-resolution via a holistic attention network. In:
  European conference on computer vision. pp. 191--207. Springer (2020)

\bibitem{vvc}
Ohm, J.R., Sullivan, G.J.: Versatile video coding–towards the next generation
  of video compression. Picture Coding Symposium  (2018)

\bibitem{bam}
Park, J., Woo, S., Lee, J.Y., Kweon, I.S.: Bam: Bottleneck attention module.
  arXiv preprint arXiv:1807.06514  (2018)

\bibitem{JPEG2000}
Rabbani, M., Joshi, R.: An overview of the jpeg 2000 still image compression
  standard. Signal processing: Image communication  \textbf{17}(1),  3--48
  (2002)

\bibitem{swish}
Ramachandran, P., Zoph, B., Le, Q.V.: Searching for activation functions. arXiv
  preprint arXiv:1710.05941  (2017)

\bibitem{hevc}
Sullivan, G.J., Ohm, J.R., Han, W.J., Wiegand, T.: Overview of the high
  efficiency video coding (hevc) standard. IEEE Transactions on Circuits and
  Systems for Video Technology  \textbf{22}(12),  1649--1668 (2012).
  \doi{10.1109/TCSVT.2012.2221191}

\bibitem{toderici2015variable}
Toderici, G., O'Malley, S.M., Hwang, S.J., Vincent, D., Minnen, D., Baluja, S.,
  Covell, M., Sukthankar, R.: Variable rate image compression with recurrent
  neural networks. arXiv preprint arXiv:1511.06085  (2015)

\bibitem{toderici2017full}
Toderici, G., Vincent, D., Johnston, N., Jin~Hwang, S., Minnen, D., Shor, J.,
  Covell, M.: Full resolution image compression with recurrent neural networks.
  In: Proceedings of the IEEE conference on Computer Vision and Pattern
  Recognition. pp. 5306--5314 (2017)

\bibitem{JPEG}
Wallace, G.K.: The jpeg still picture compression standard. IEEE transactions
  on consumer electronics  \textbf{38}(1),  xviii--xxxiv (1992)

\bibitem{msssim}
Wang, Z., Simoncelli, E., Bovik, A.: Multiscale structural similarity for image
  quality assessment. In: The Thrity-Seventh Asilomar Conference on Signals,
  Systems Computers, 2003. vol.~2, pp. 1398--1402 Vol.2 (2003).
  \doi{10.1109/ACSSC.2003.1292216}

\bibitem{cbam}
Woo, S., Park, J., Lee, J.Y., Kweon, I.S.: Cbam: Convolutional block attention
  module. In: Proceedings of the European conference on computer vision (ECCV).
  pp. 3--19 (2018)

\bibitem{vimeo}
Xue, T., Chen, B., Wu, J., Wei, D., Freeman, W.T.: Video enhancement with
  task-oriented flow. International Journal of Computer Vision (IJCV)
  \textbf{127}(8),  1106--1125 (2019)

\bibitem{res_ch_attn}
Zhang, Y., Li, K., Li, K., Wang, L., Zhong, B., Fu, Y.: Image super-resolution
  using very deep residual channel attention networks. In: Proceedings of the
  European conference on computer vision (ECCV). pp. 286--301 (2018)

\bibitem{end-to-end-attention}
Zhou, L., Sun, Z., Wu, X., Wu, J.: End-to-end optimized image compression with
  attention mechanism. In: CVPR workshops. p.~0 (2019)

\end{thebibliography}

\end{document}